\documentclass[letterpaper, 10pt, conference]{ieeeconf}
\usepackage{enumerate}
\usepackage{bm}

\usepackage{ifthen}
\usepackage{lineno}
\usepackage{amssymb}
\usepackage{amsmath}
\usepackage{adjustbox}
\usepackage{epsfig}
\usepackage{graphicx}
\usepackage{graphics}
\usepackage{float}
\usepackage{bbm}
\usepackage{color}

\usepackage{soul}

\newcommand{\eop} {\hfill{$\blacksquare$}}
\newcommand{\Cmnt}[1] {}
\newtheorem{theorem}{Theorem}

\newtheorem{lemma}{Lemma}

\definecolor{darkred}{rgb}{1, 0.1, 0.3}
\definecolor{darkblue}{rgb}{0.1, 0.1, 1}
\definecolor{darkgreen}{rgb}{0,0.6,0.5}

\newcommand {\mm}[1] {\ifmmode{#1}\else{\mbox{\(#1\)}}\fi}

\newcommand{\AsymCmnt}[1]{}   

\newcommand{\TR}[2]{#2}  

\renewcommand{\a}{{\bf a}}

\newcommand{\hide}[1]{}

\newboolean{showcomments}
\setboolean{showcomments}{true}
\newcommand{\jk}[1]{\ifthenelse{\boolean{showcomments}} {\textcolor{red}{(JK says: #1)}} {} }
\renewcommand{\ss}[1]{\ifthenelse{\boolean{showcomments}} {\textcolor{red}{(SS says: #1)}} {} }
\newcommand{\vk}[1]{\ifthenelse{\boolean{showcomments}} {\textcolor{red}{(VK says: #1)}} {} }

\newcommand{\fun}{{\cal Z}}

\makeatletter
\let\NAT@parse\undefined
\makeatother
\usepackage{hyperref}

\begin{document}

\title{Social Optimal Freshness in Multi-Source, Multi-Channel Systems via MDP} 
\author{
$\text{ Shiksha Singhal, Veeraruna Kavitha}$ and $\text{Vidya Shankar}$  \\
$\text{ IEOR}$, Indian Institute of Technology Bombay, India
}




\maketitle
\setcounter{page}{1}

\begin{abstract}
Many systems necessitate frequent and consistent updates of a specific information.
Often this information is updated regularly, where an old packet becomes completely obsolete in the presence of a new packet. In this context, we consider a system with multiple sources, each equipped with a storage buffer of size one, communicating to a common destination via $d$ orthogonal channels. 
In each slot, the packets arrive at each source with certain probability  and occupy the buffer (by discarding the old packet if any), and each transfer (to the destination) is successful with certain other probability. 
Thus in any slot, there are two (Age of Information) AoI-measures   for each source: one corresponding to the information at the source itself and the other corresponding to the information of the same source available at the destination; some sources may not even have the packet to transmit. 

The aim of the controller at the destination is to maintain the freshness of information of all the sources, to the best extent possible -- it aims to design an optimal scheduling policy that assigns in each slot,  a subset of sources with packets (at maximum $d$)  for transmission.
This is achieved using 
an appropriate Markov Decision Process (MDP) framework,  where the objective function is the  sum of Average AoIs (AAoI)  of all the sources. We derive a very simple stationary policy that is $\epsilon$-optimal -- in any slot, order the sources with packets in the decreasing order of the  differences  in AoI at the destination and the source and choose the top sources for transmission. With moderate number of sources $(<30)$, the AAoI reduces in the range of $30-90\%$.
\end{abstract}

\section{Introduction}
With the advent of new technology and next generation networks that support smart applications, the need to  continuously update information at centralised location from various sources becomes increasingly imperative; 
for example, Internet of Things (IoT), smart homes, environmental monitoring systems,  on-the-road communication retrieval systems etc.  
The sources of information are  required to transmit periodic status updates to their intended destinations (see \cite{bedewy2016optimizing}-\cite{corke2010environmental}).    
A critical requirement for these services is ensuring that the information provided by the sources remains up-to-date at the destination, to the best extent possible; the quality of freshness of information is measured using  ``age of information'' (AoI) which is the time elapsed since the time of generation of the latest available information \cite{kaul2011minimizing}. However, this task is  constrained by limited  resources and the requirement  coming from multiple sources. Thus we consider designing an optimal scheduling policy that optimises  the sum of average age of information (AAoI) of all the sources. 

Further, often in the systems that require regular updates of the same information, the old packet becomes obsolete once a new packet is available (\cite{kavitha2021controlling}). Thus it is more appropriate to consider systems with at maximum one buffer storage, leading to lossy systems.

Motivated by the above factors, we consider a time-slotted system consisting of
multiple sources (with single storage) that communicate to a  common destination via multiple orthogonal channels. 
In each slot, the packets arrive at each source with certain probability, and each transfer (to the destination) is successful with certain other probability.  
Thus in any slot, there are two AoI-measures   for each source: one corresponding to the information at the source itself and the other corresponding to the information of the same source available at the destination; some sources may not even have the packet to transmit (this happens when a source does not receive a new packet after its last successful transfer). At any time slot, these two AoI-measures corresponding to all the sources represent the state of the system.
The goal is to design an optimal scheduling policy which determines the  subset of sources for transfer in any time slot and which minimises the sum of the AAoI (at the destination) corresponding to all the sources. 


\noindent
\textbf{Related work:} The problem of minimising the age of information in such systems has been studied in 
\cite{sun2017update}-\cite{yates2012real} 
which focus on push-based communication where the sources decide when they want to send an update to the destination, and hence answer questions of optimal packet generation times. 
On the other hand, \cite{tripathi2017age},\cite{kadota2018scheduling},\cite{hsu2017age} and \cite{kadota2019minimizing} focus on systems which implement pull-based communication, where the destination asks for data from the sources. 
Ours is a pull-based communication but with random packet arrivals to the sources and with some sources not having packets.

In \cite{tripathi2017age}, authors consider multiple independent sources providing status updates to a single destination via multiple orthogonal channels. The question here is similar: one needs to  optimally choose a subset of sources to transmit, in each time slot. They assume the knowledge of (binary) channel conditions  and  
choose for each channel one source among  those that can communicate with the given channel in the given time slot. Further the sources always have information to transmit. In contrast, in our work the random channel conditions are unknown and we have more uncertainty in terms of packet availability at sources --- it is not  realistic to assume that the measurements are always available in all the time slots (measurement errors, transmission problems from the point of generation to the source itself, etc.), e.g., as in stock updates, sensor measurement, IoT,  etc., which leads to a possibly  stale (one or more slot old information) information even at the sources. 
Thus the policies in \cite{tripathi2017age} consider only the AoIs at destination while we consider both sets of ages, including the ages at the sources. 


The authors in \cite{kadota2019minimizing} and \cite{hsu2017age} consider an infinite time horizon problem and find optimal stationary randomised policies that are blind to the state of the system,  i.e., without taking into account the ages of the packets at sources and the queue lengths (systems are not lossy) etc. The former work does not even  consider the age of information at destination, while the latter work assumes perfect transmissions, i.e.,  the transmission probability equals one.


The study in \cite{kadota2018scheduling} considers a framed-slot structure where the packets are generated in the beginning of each frame for each source and 
the scheduling policy allocates at maximum one source in each slot of the frame, based on the success of the previous transmission attempts of the same frame. Hence like in \cite{tripathi2017age}, the  study in \cite{kadota2018scheduling} also assumes the availability of packet at each source and in each frame and their policy is also similar; in each slot of the frame, select the source with highest age   at destination and  switching to a new source only when the packet is transmitted in the previous slot. 
This paper considers many more interesting aspects however, does not consider uncertainty related to packet availability at sources. 



In contrast to the above strands of literature, we consider a system with unreliable and unknown channels (scheduling is blind to the channel conditions and hence scheduled transmission is successful with  probability $p$) and uncertainty in packet availability at sources. We however observe the AoIs at all the sources and destination for optimal source selection.  
Towards this, we derive an $\epsilon$-optimal scheduling policy and compare its performance with two policies inspired by the existing policies in literature; as already mentioned, none of the existing algorithms work under our assumptions (mainly uncertain packet availability at sources) and hence, we adapt them to our scenario and then compare with the proposed policy. In particular, we compare the proposed policy with: i) round-robin (RR) policy which chooses a subset of  sources one after the other irrespective of the instantaneous ages  and, ii) 
a partial information (PI) policy that chooses a subset of sources for transmission in any slot only based on instantaneous ages at the destination.   

\noindent 
\textbf{Contributions: }The main contributions of this work are:
\begin{enumerate}[(a)]
    \item We formulate the problem as an appropriate Markov Decision Process (MDP) and derive an $\epsilon$-optimal policy. Its  performance approaches the performance of the optimal policy as $p^2$ reduces to zero (Theorem \ref{thm_eps_optimal}). The policy is defined by a simple stationary rule, i.e., in any slot, order the sources with packets in the decreasing order of the  differences  in AoI at the destination and the source and choose the top sources for transmission, 
    \item We  demonstrate significant improvement (even up to $90\%$) with the moderate number of sources (Section \ref{sec_sim}). More interestingly, the performance improvement is significant  for all values of $p$ (Figures \ref{fig_N_q1} and \ref{fig_N_q2}) and,
    \item The performance of our policy starts matching with the existing policies when the number of sources is large (approximately $>100$).
\end{enumerate}

\section{Problem Definition}
Consider a system with a set $\mathcal{N} = \{1,\cdots,N\}$ of $N$ sources, sending regular updates (packets) of a certain information 
to a common destination via $d$ orthogonal channels. We use a time-slotted system with $T$ number of slots, where in each time slot of length $\tau$, each channel can be used by at most one source to transfer its packet to the destination. In every time slot, the packets arrive at any source $n$ according to a Bernoulli process with probability $q_n$ and a successful packet transfer to the destination happens with probability $p$ for any source. All these events are independent of each other. The packets are identical but the transfer times may vary based on the random channel conditions -- note here that the packet transfer times are geometric with parameter $p$ for each channel.   Each source has its own buffer with the storage capacity of one. \textit{It is sufficient to consider storage capacity of one at every source, as the old packet becomes obsolete once a new packet arrives}.
Our focus in this work is on measures related to the freshness of information available at the destination related to all the sources. Towards this, we measure the quality of information using  a metric, called Age of Information (AoI) \cite{kaul2011minimizing}.

\vspace{1.5mm}
\noindent
\textbf{Age of Information (AoI): } The age  $(H_n)$ of information  of  source $n$ at  destination and at time-slot $t$ is defined as the time elapsed since the last received packet at destination was generated, i.e.,
\begin{equation}
\label{eqn_AoI}
    H_n(t) := t - r_n(t),
\end{equation}
where $r_n(t)$ is the time at which the last successfully received packet (at destination) before time $t$, is generated at source $n$. 

\noindent
\textbf{Average Age of Information (AAoI): }For any $n \in \mathcal{N}$, it is defined as below,
\begin{equation}
    {\bar h}_n := \frac{\sum_{t=1}^T  H_n(t) }{T},
    \label{eqn_aaoi}
\end{equation}
where $T$ is the required time horizon.
As already mentioned, we consider freshness of information in a lossy system, and our aim is to find an optimal source selection policy (for transmission in each slot) which minimises the sum of AAoI at the destination, from all the sources.

\begin{figure}[h]
    \centering   \includegraphics[trim = {0cm 1.5cm 5cm 1cm}, clip, scale = 0.2]{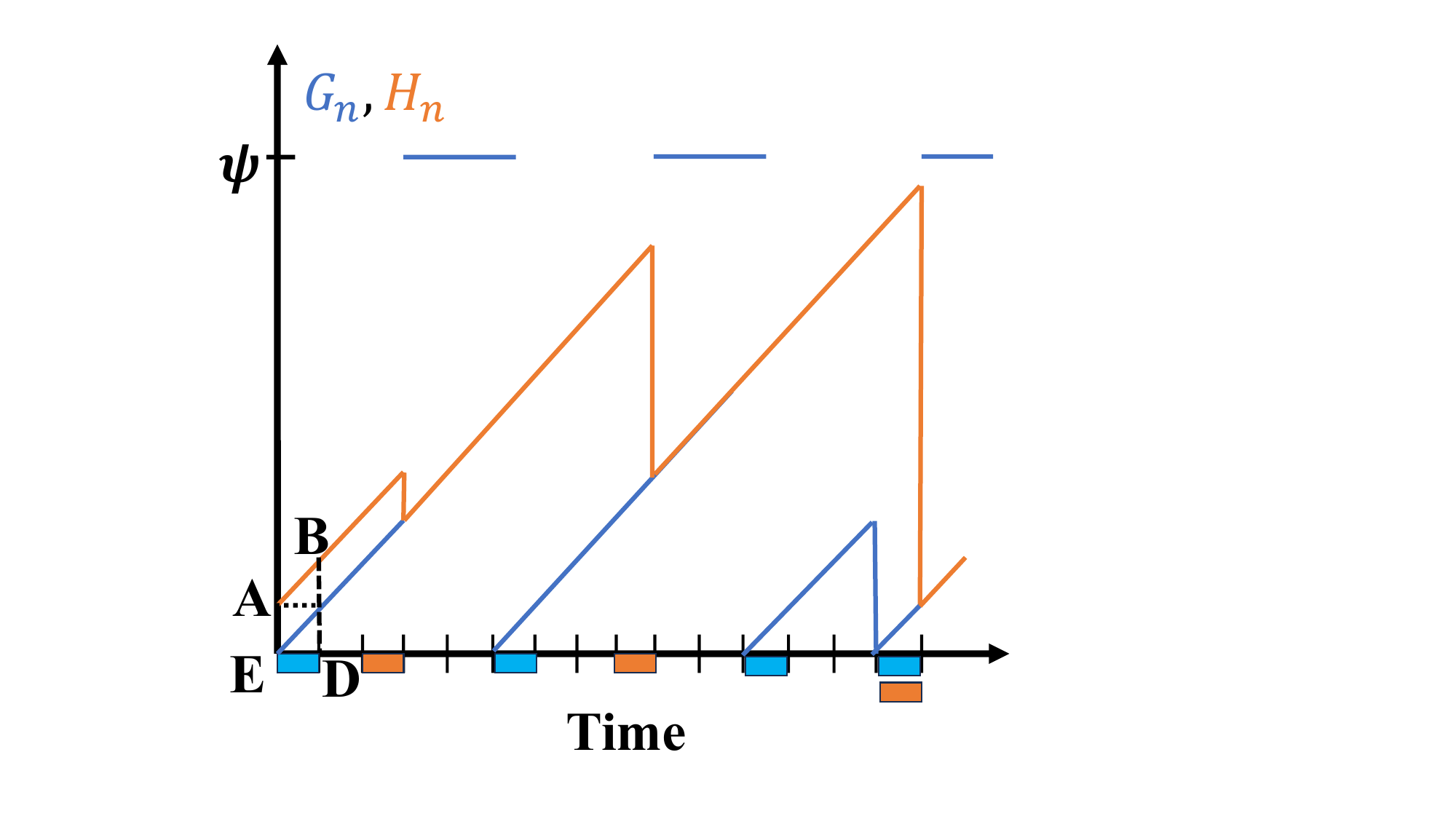}
   \caption{Evolution of age at source and destination with time where blue and orange rectangles represent packet arrival to source $n$ and successful packet transfer from source $n$ to destination, respectively.}   \label{fig_evolution_age}
\end{figure}

In this context, the state of the system at any time $t$ can be represented by $X(t) = ({\bf G}(t),{\bf H}(t))$ which is made up of two vectors with ${\bf H}(t)= (H_1(t),\cdots,H_N(t))$ as defined in \eqref{eqn_AoI} while ${\bf G}(t)= (G_1(t),\cdots,G_N(t))$ corresponds to the age of information at all the sources (see Figure \ref{fig_evolution_age}). 
For example, if $g_n(t) =k$ then the latest packet waiting at source $n$ is generated in $({t-k})^{th}$ slot. 

Another important point here is to observe that the old waiting packet is replaced with the new packet, if there is any arrival in any slot. 
 Further, the age of any source $n$ at time $t$ is replaced with symbol $\psi$, if it has no packet; this can happen if its latest packet has been transmitted and there was no new packet arrival after that. \textit{It is easy to see that $G_n < H_n$ whenever source $n$ has a packet.}

 Note that the decision is simplified to selecting a  subset of  sources at the beginning of each time slot, instead of having to choose whether to continue with the current packet or choose a new packet (from the same/different source) as the transfer times are geometric and exhibit memory-less property. We model this as a finite horizon Markov Decision Process (MDP) whose ingredients we describe in the next section.

 The aim of this paper is to derive optimal source scheduling policy which minimises the sum of the AAoI from all the sources at the destination (see \eqref{eqn_aaoi}), i.e., 
 \begin{equation}
 \label{eqn_obj}
    \min \sum_{n \in \mathcal{N}} {\bar h}_n.
 \end{equation}

\section{ MDP Formulation}
\label{sec_MDP}
In this section, we describe the  MDP formulation and its ingredients.

\vspace{1mm}
\noindent
\textbf{Decision Epochs:} The beginning of each time slot is considered as a decision epoch. We have $1,\cdots,T$ decision epochs with $T < \infty$. 

\vspace{2mm}
\noindent
\textbf{States: }As already defined, the state of the system at any time $t$, $X(t) = (\textbf{G}(t),\textbf{H}(t))$ which contains AoIs at all sources and destination  at the beginning of time slot $t$. The state space is, 
 $$\mathcal{X} = \{x = ({\bf g},{\bf h}): g_i = \psi \text{ or } g_i < h_i \}.$$
\TR{ Let $\mathcal{X}_\psi = \{x \in {\cal X}: g_n = \psi \text{ for all } n\}$ be the subset of states at which no source has a packet to transmit.}{}

\vspace{2mm}
\noindent
\textbf{Actions:} 
The action at every decision epoch is to choose a subset of sources ${\bf a}$, whose packet is to transferred. Let $N_x:= \sum_n 1_{\{g_n \ne \psi\}}$  be the number of sources that have packets to transmit. Then, the state dependent action space,
\begin{eqnarray}
   \mathcal{A}_{x} =  2_{N_x^d}^{S_x} \text{ with } S_x:= \{ n: g_n \ne \psi  \} , N_x^d := \min \{N_x,d\},
   %
\end{eqnarray}
where $2_{N_x^d}^{S_x}$ is a $N_x^d$ sized subset of $S_x$, the set of sources with packets that can be transmitted. There are some observations related to this definition.
Observe that there is only one action and ${\cal A}_x = \{S_x\}$  for all the states with $|S_x| = N^d_x \le d $ and that no transmission is attempted if $N_x^d = 0$. Further, when $N_x > d$, any action ${\bf a} \subset S_x$ with $|{\bf a}| = d$. In other words, we consider work conserving policies that facilitate transfer of all the available packets, however obviously constrained by capacity $d$. 




\vspace{2mm}
\noindent
\textbf{Cost: }Towards optimising \eqref{eqn_obj}, from \eqref{eqn_aaoi} and \eqref{eqn_obj}, the appropriate instantaneous cost at time $t$ when the state is $x \in \mathcal{X}$ and action $\a \in \mathcal{A}_x$ is chosen  is given by, 
$$
c(x,\a)= \mathbb{E} \left[ \sum_{n =1}^N h_n | x,\a \right] \text{ for all } \a \in \mathcal{A}_x \text{ and all } x.
$$ 
The above conditional expectation is the sum of $N$ terms each of which is area under the age curve corresponding to one source (area under trapezium ABDE in Figure \ref{fig_evolution_age} is one such area corresponding to source $n$)  during a given time slot. Thus, 
$$
c(x,\a)= \sum_{n=1}^N \left(h_n \tau + 0.5\tau^2  \right) \text{ for all } \a \in \mathcal{A}_x \text{ and all } x.
$$ 
Now optimising using the above instantaneous cost is equivalent to optimising using instantaneous costs, which equals
$
   c(x,\a)= \sum_{n=1}^N h_n \tau \text{ for all } \a \in \mathcal{A}_x \text{ and all } x,
$ 
as $\tau$ is an arbitrary constant. Thus, we set  
\begin{equation}
   c(x,\a)= \sum_{n=1}^N h_n \text{ for all } \a \in \mathcal{A}_x \text{ and all } x.
   \label{Eqn_cost}
\end{equation}

\vspace{2mm}
\noindent
\textbf{Transition probabilities: }The system evolves to a new state $X'$ when action $\a \in \mathcal{A}_x$ is chosen in state $x \in \mathcal{X}$ at any decision epoch. This transition  depends on the transfer status (if any, i.e., if $N_x^d > 0$) and packet arrival status, both in the previous slot. 
Let  $\mathcal{C} \subset \mathcal{N}$ be the set of sources which generated new packets  in the duration between previous and current decision epoch.  Let $X(t-1)=x$  and ${\bf A}({t-1}) = \a.$

Now, if the packet of  sources $W \subset \a$ is successfully transferred then the state evolves as below: for any $n$,
\begin{eqnarray}
\label{eqn_h_evol}
H_{n}' & = & 
\begin{cases}
h_n + 1 &
 \text{for all } n \notin W, 
\\
g_{n} + 1 & \text{else, i.e., if } n \in W.
\end{cases} \\
\label{eqn_g_evol}
G_{n}' & = & 
\begin{cases}
g_{n} + 1 & \text{for all }  n \notin  \mathcal{C}\text{ and } n \notin W, \\
0 & \text{for all }  n \in \mathcal{C} , \\
\psi & \text{else, i.e., if } n \notin \mathcal{C} \text{ and }n \in W.
\end{cases} 
\end{eqnarray} 
Observe that the age of packet at source $a$, $G'_{a}$ is set to $\psi$ when it does not receive a new packet, as the latest packet with it is just transferred. The transition of the above type (specified by  $W$ and ${\cal C}$) happens with probability,

\vspace{-4mm}
{\small \begin{equation}
   \mathbb{P}(X(t) = X'|x,\a) =  p^{|W|}(1-p)^{|\a|-|W|} \prod_{n \in \mathcal{N} \setminus \mathcal{C}  }(1-q_n) \prod_{n \in  \mathcal{C}  }  q_n.
   \label{eqn_trans_prob}
\end{equation}}

\TR{On the other hand, if the packet transfer of source $a$ fails, then the state evolves to the following,
\begin{eqnarray*}
H_{n}' & = & H_{n}+1 \text{ for all }  n \in \mathcal{N}, \\
G_{n}' & = & 
\begin{cases}
G_{n} + 1 & \text{for all }  n \in \mathcal{N} \setminus \mathcal{C}, \\
0 & \text{for all } n \in \mathcal{C}, \\
\end{cases} \\
\end{eqnarray*}
with transition probability,
\begin{equation*}
\mathbb{P}(X(t) = X'|x,a) = (1-p) \prod_{n \in \mathcal{N} \setminus \mathcal{C}  }(1-q_n) \prod_{n \in  \mathcal{C}  }  q_n.
\end{equation*}}{}
Thus with all the  ingredients defined, the optimisation of the term in \eqref{eqn_obj} is equivalent to solving the following MDP where, $$
V^*(x)  = \min_{\pi = (d_1,\cdots,d_{T-1})} \mathbb{E}_{\pi, x}  \left[ \sum_{t=1}^T c(X(t),{ \bf A}(t))\right].
$$

\section{Near Optimal Policy}
\label{sec_eps_optimal}

\TR{Consider a set of initial conditions,

\vspace{-4mm}
{ \small \begin{equation}
    {\cal X}_0(M) := \{x \in {\cal{X}}: h_n \le M\} \text{ for all }n \in \mathcal{N} \text{ and } 0 \le M < \infty. \nonumber
\end{equation}}}{}
\TR{Further, let  ${\cal X}_\psi \subset {\cal X}$ be the set of states with $G_n = \psi$ for all $n$.}{} 
\noindent
The aim of this section is to derive near optimal or $\epsilon$-optimal policy. It is well known in the MDP literature that the optimal policy can be non-stationary (i.e., the decision rules are different across time slots) for finite horizon problems \cite{putterman}. However, interestingly our $\epsilon$-optimal policy turns out to be stationary. 
In particular, we consider the following special stationary policy constructed using the differences between AoIs at sources and destinations as defined below:

\vspace{-4mm}
{\small \begin{eqnarray}
\begin{aligned}
   \pi^\Delta = (d_1^\Delta,\cdots,d_{T-1}^\Delta) \text{ where } d_t^\Delta = d^\Delta \text{ for all }t, \text{ and }\\
   d^\Delta(x) 
 = \arg \min_{{\bf a} \in {\cal A}_x  } \sum_{i =1}^{ N_x^d } ( g_{a_i} - h_{a_i} ),   
 %
\end{aligned}
\label{Eqn_eps_opt_policy}
\end{eqnarray}}as  already explained (in Section \ref{sec_MDP}) there is only one action when $N_x \le d$.
Next, we present our main result which upper bounds the performance of $\pi^\Delta$ policy with respect to the optimal value (proof in Appendix).

 \begin{theorem}
{ \it 
There exists a sequence of non-negative functions $\{\fun_{t}(x)\}$ for each $t \le T$ such that 
the stage-wise optimal value function(s) $\{V^*_t (x)\}$ are  related to  the corresponding value function(s) $\{V^{\Delta}_{t}(x)\}$ under policy $\pi^\Delta$ of \eqref{Eqn_eps_opt_policy}   as below:
\begin{eqnarray}
\label{eqn_diff}%
V^{\Delta}_{t}(x) - V^*_{t}(x)   &=&  p \cdot p_d \fun_{t}(x) \mbox{ for each }  t, x, \mbox{ with, }
  \\
  && \hspace{-32mm}
   p_d \ := \ 1-(1-p)^d = p C_p, \  C_p = \left( \sum_{l=0}^{d-1} (-1)^{l} \binom{d}{l+1} p^{l}\right).   \nonumber 
\end{eqnarray}
Further there exist two constants $\{D_1(t)\}$ and $\{D_2(t)\}$ independent of $p$ such that,

\vspace{-3mm}
{\small 
\begin{eqnarray}
\label{eqn_Z_bd_thm}
|\fun_t(x)| & \le & D_1(T-t) ||x||_\infty + D_2(T-t) \mbox{ with }  \\  
|| x||_\infty  & := & \max_{n \in \mathcal{N}}\{ h_n+1 \}. \nonumber
\end{eqnarray} }
}
\label{thm_eps_optimal}
\end{theorem}
\noindent
\textbf{Remarks: }From the above theorem  (see \eqref{eqn_diff}-\eqref{eqn_Z_bd_thm}) the difference,
$$  
V^{\Delta} (x) - V^* (x)   \le  p \cdot p_d  \left (D_2 (1) + ||x||_\infty D_1 (1)   \right ) \mbox{ for each }   x, 
$$
where constants $D_1(1), D_2(1)$ are independent of $p$. Thus clearly as $p \to 0$, we have $p \cdot p_d = p^2 C_p  \to 0$ and thus the objective function  $ V^{\Delta} (x) $ evaluated under stationary policy $\pi^\Delta$
of \eqref{Eqn_eps_opt_policy}   approaches the value function $ V^* (x) $. In fact this approach is uniform in all $x \in {\cal X}_B $ for any $B < \infty$, where  ${\cal X}_B := \{ x : g_n = \psi \text{ or }  g_n < h_n \mbox{ and } h_n \le B \ \forall \ n  \} $.

\textit{Thus $\pi^\Delta$ is $\epsilon$-optimal} when initial condition belongs to ${\cal X}_B$ and when $p \sqrt{C_p} < \sqrt{  \epsilon / (D_1(1) (B+1) + D_2 (1)  )   } $. This establishes near-optimality of   $\pi^\Delta$ when $p^2 $  is sufficiently small (observe $C_p \le d$ for all $p$). 

Observe that the above approximation is good as $p^2 \to 0$ and this already suggests that even for moderate $p$ (and for which $p^2$ is negligible) one can have a good approximation. In fact, more interestingly we observe via simulations in the next section that the $\pi^\Delta$ policy performs significantly well even for values of $p$ close to  one, in comparison with the policies known in literature.

\medskip

\TR{\begin{enumerate}[(i)]

    \item In fact in the proof (see \eqref{eqn_val_k}- \eqref{eqn_Z_diff}) we can see that the function ${\cal Z}_t (x)$ as well as $V^*_t(x)$ both grow with depth $(T-t)$ in a similar manner (both are  polynomial  in depth of same degree). Thus one can anticipate a good approximation when $p^2$ is small. Further observe that $p^2$ can be much smaller than $p$ and hence can anticipate a good approximation 
    even for $p$ sufficiently big, and we observe the same in the Section \ref{sec_sim} with numerical examples. 
    
   There exists  bounded function $\{\mathcal{F}_t(x)\}$ for $t \le T$ and the constant $K(t)$, such that using \eqref{eqn_val_eps}-\eqref{eqn_common}, the value function(s) under policy $\pi^{\Delta}$  of \eqref{Eqn_eps_opt_policy} is given by:

\vspace{-3mm}

{\small \begin{eqnarray*}
 \hspace{-10mm}  V^{ \Delta}_{T-k}(x)  & = & (k+1)\sum_{n \in \mathcal{N}}h_n + K(k) \nonumber   + p\mathcal{F}_{T-k}(x) \\
  \hspace{-20mm}  && + kp \min_{\a \in \mathcal{A}_x} \sum_{i=1}^{N_x^d} \ (g_{a_i}-h_{a_i})    \label{Eqn_eps_value_func}
\end{eqnarray*}}

for any $0 \le k \le (T-1)$ with finite constants $\{C_1(t)\}$ and $\{C_2(t)\}$, the function $|{\cal F}_t(x)| \le C_1(t)||x||_\infty+C_2(t)$,  $K(k) = kN+K(k-1)$, and $K(0) = 0$.
 
\end{enumerate}}{}

\section{Simulations}
\label{sec_sim}

In this section, we compare the performance
of our proposed policy with the  policies in the existing literature. 
To the best of our knowledge, it appears that there is no work in the literature that considers our scenario, i.e., unreliability at packet generation and transfer; further the decision maker does not have access to the channel conditions.

The authors in \cite{tripathi2017age} consider generic binary channels (for example, a kind of Markovian channel) but assume  the availability of the packets at all the sources, in each time slot; they also assume the knowledge of the channel conditions before taking a decision. Their policy is to select a subset of sources (among the sources with good channel conditions) with the highest AoIs at destination for transfer, in each time slot.

On the other hand, \cite{kadota2018scheduling} considers a framed slotted structure, with each source having a packet at the beginning of the frame. They propose a modified Robin Round (greedy) policy, which order the sources in the decreasing order of AoIs at the destination in the first frame. Henceafter, they follow the Round Robin policy, i.e., chooses sources one after the order, with switching only once the packet is transferred. The specific choice in the first frame  and the Round Robin  henceafter ensures that the source selected in any time slot is the one with the highest AoI at the destination.

We compare our $\epsilon$-optimal policy  \eqref{Eqn_eps_opt_policy} with the one obtained by adapting the policies in \cite{tripathi2017age} and \cite{kadota2018scheduling} to our scenario, i.e., to the case with unreliable packet generation --- basically, at any time instance we choose a subset of sources based on the order of the AoIs at the destination among the sources with packets to transfer. Since this policy does not consider the AoI at sources, we refer it as partial information (PI) policy.

We also compare our policy (referred to as $\epsilon$-O policy) with a complete blind policy, which chooses sources for transfer one after the other, irrespective of the AoIs and the transfer status of the packets. We refer it as
Round Robin (RR) policy.

\begin{figure}[h]
   \begin{minipage}{.24\textwidth}
    \includegraphics[trim = {0cm 5.5cm 0cm 7cm}, clip, scale = 0.2]{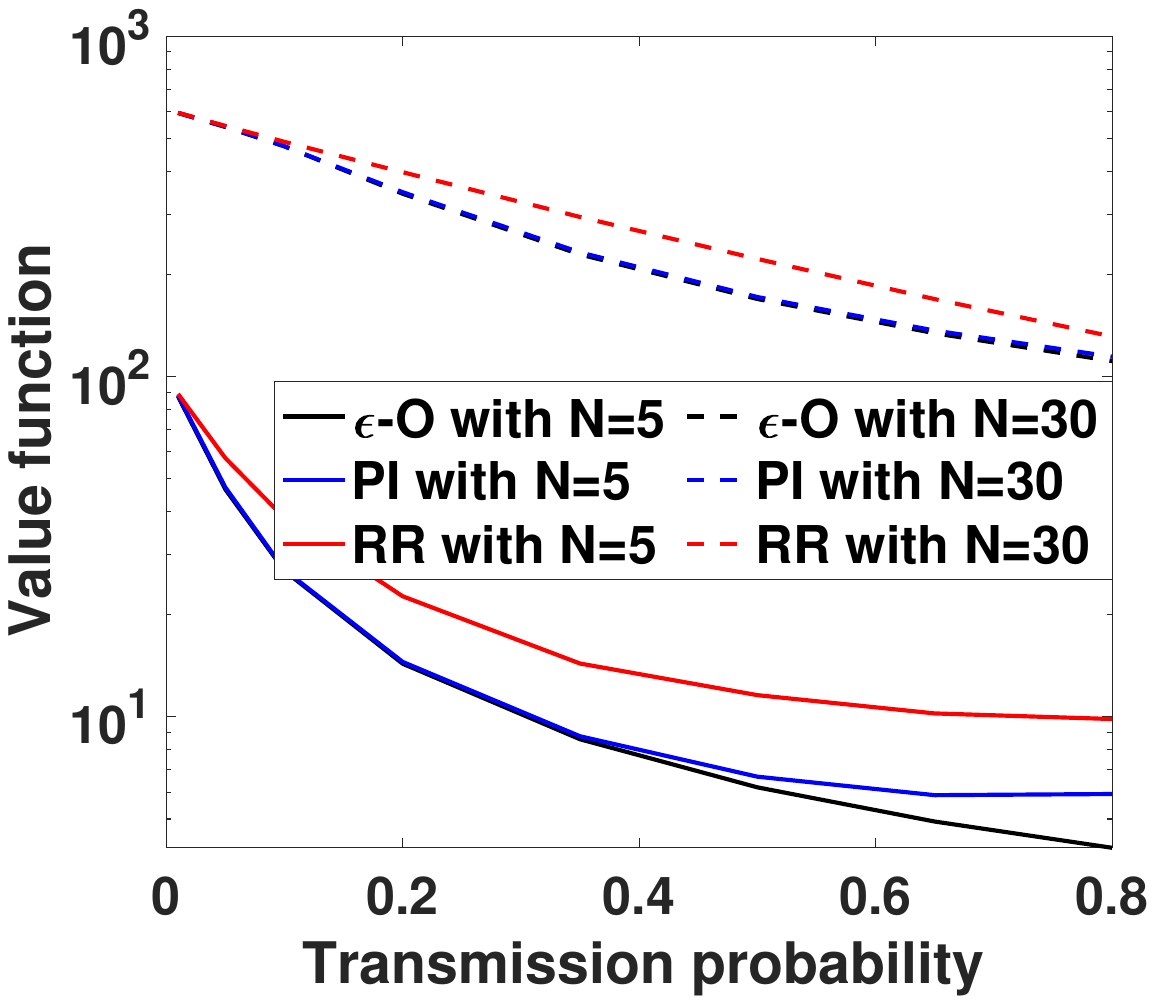}
    \end{minipage}
    \begin{minipage}{.24\textwidth}
    \includegraphics[trim = {0cm 6.5cm 1cm 7.4cm}, clip, scale = 0.2]{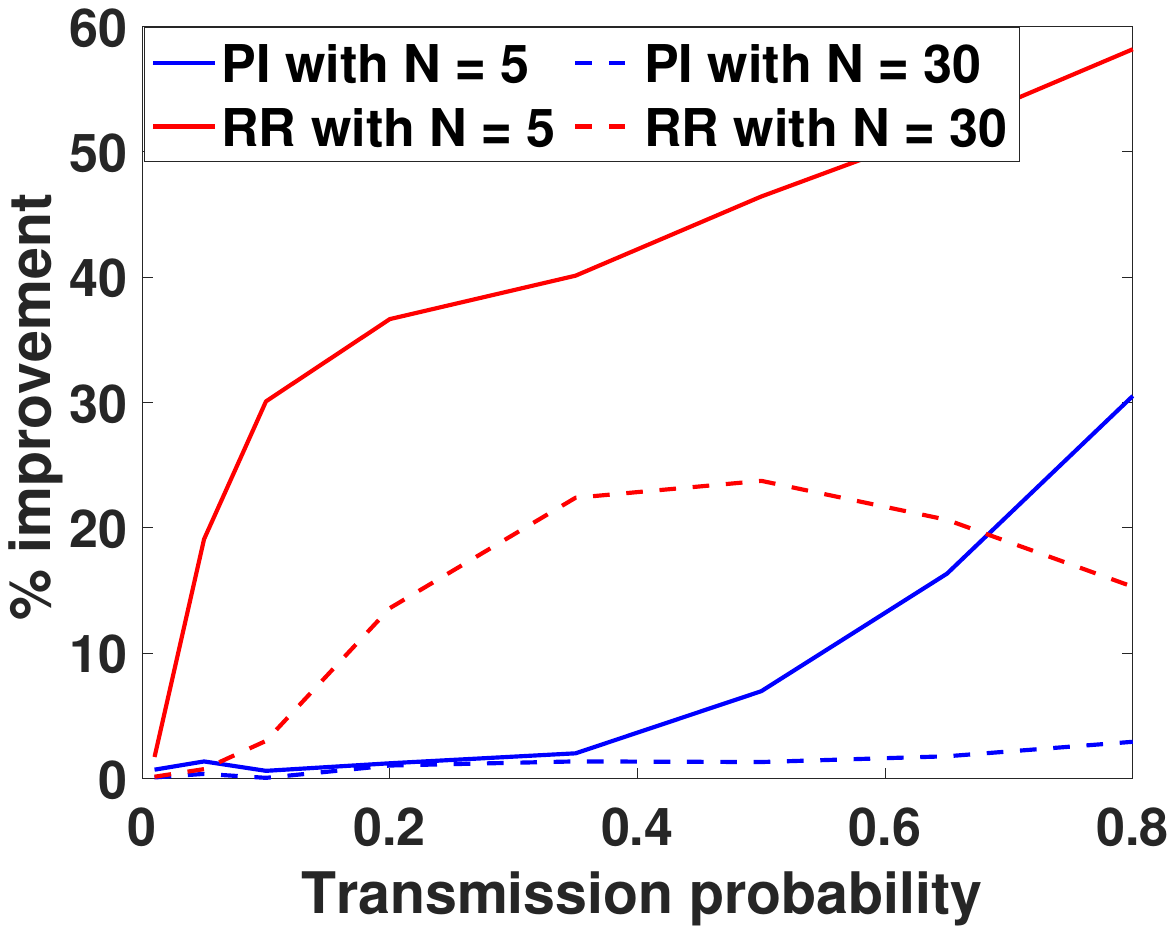}
    \end{minipage}
     \caption{{\bf Packet arrivals, $q_n = 0.5$ $\forall \ n$:} Value function (on log scale) in left sub-figure  and percentage improvement in the right figure versus transmission probability $p$ with $d=1$   \label{fig_N_q1}}
\end{figure}

\begin{figure}[h]
   \begin{minipage}{.24\textwidth}
    \includegraphics[trim = {0cm 5cm 0cm 7cm}, clip, scale = 0.2]{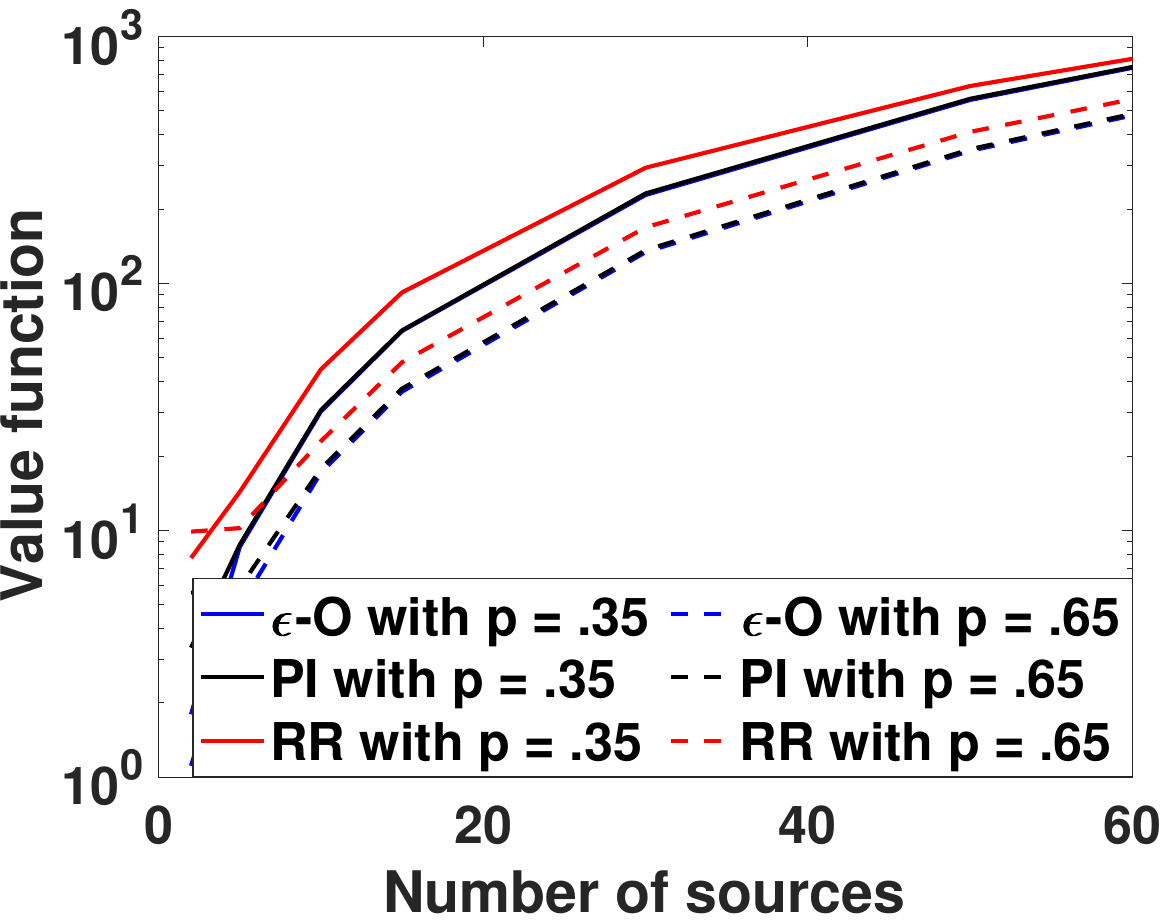}
    \end{minipage}
    \begin{minipage}{.24\textwidth}
    \includegraphics[trim = {0cm 5cm 1cm 6cm}, clip, scale = 0.2]{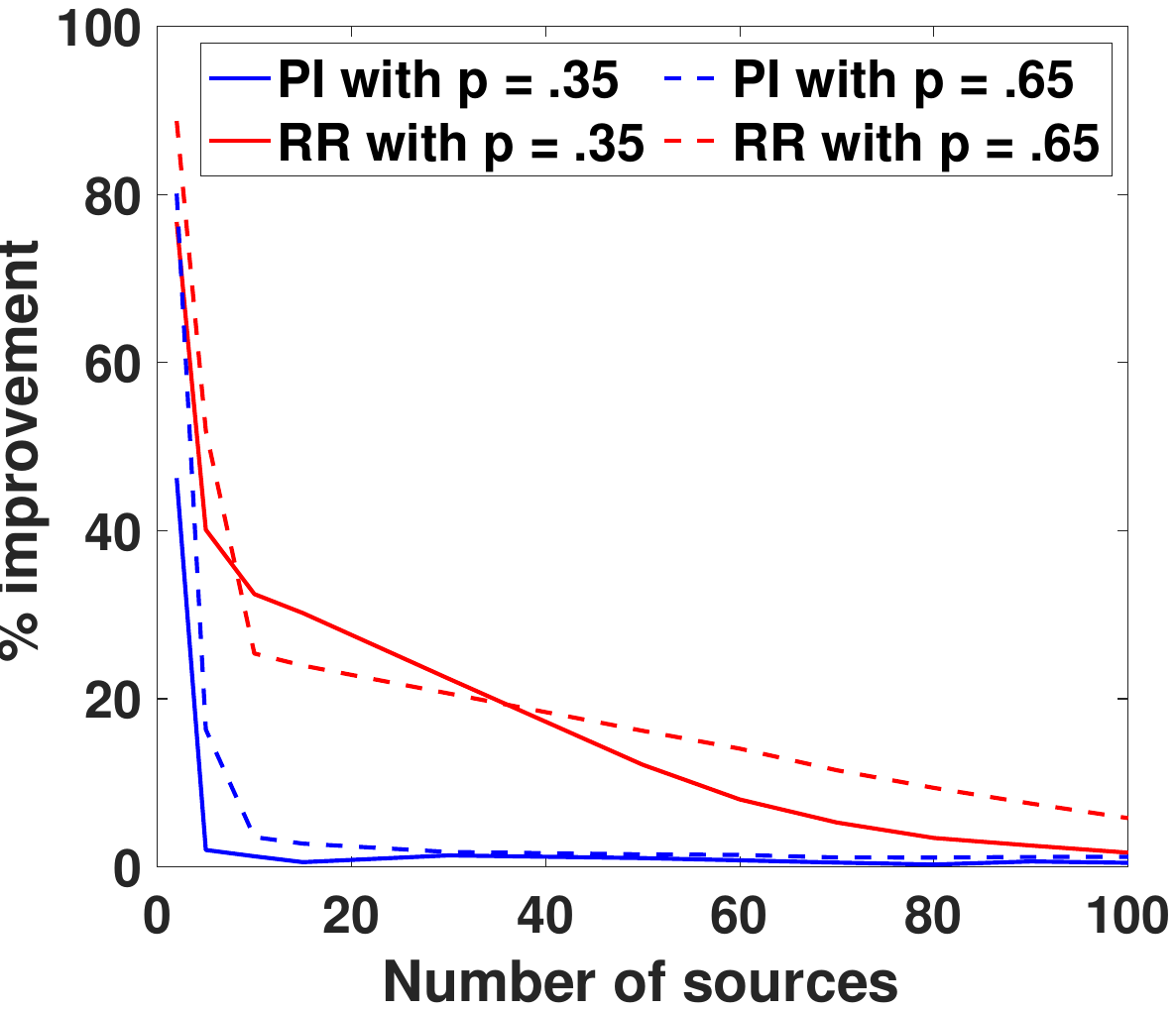}
    \end{minipage}
     \caption{{\bf Packet arrivals, $q_n = 0.5$ $\forall \ n$:} Value function (on log scale) in left sub-figure  and percentage improvement in the right figure versus number of sources $N$ with $d=1$   \label{fig_p_q1}}
\end{figure}

In Figures \ref{fig_N_q1}-\ref{fig_p_q1}, we consider an example with  packet arrival probabilities $q_n = .5$ for all sources $n \in \mathcal{N}$ and $d=1$. We consider comparison across different number of sources or different values of transmission probability $p$, in this study. One can make several observations as below,
\vspace{-5mm}
\begin{enumerate}[(a)]

  \item In Figure \ref{fig_N_q1}, we consider comparison across various $p$ for two different values of $N$, the solid lines are for $N=5$, while the dash lines are for $N=30$. Different policies are represented by different colours. 
  It is easy to see from the figures that the improvement in value function as compared to the PI policy is almost negligible for smaller values of $p$ while it is large and up to 55\% for higher values of $p$. Further, even though for $\epsilon$-O policy, theoretical guarantees have been established for  smaller values of $p$, it still performs better than RR and PI for higher values of $p$.

    \item In Figure \ref{fig_p_q1}, we consider comparison across various number of sources for two different values of $p$, the solid lines are for $p=0.65$, while the dash lines are for $p=0.35$. Different policies are again represented by different colours.  
    
    Clearly,~$\epsilon$-O policy and PI policy outperforms the RR policy (given by red curves); this is true even in Figure~\ref{fig_N_q1}. However, more interestingly, as the number of sources increases, the performance of RR policy also approaches the performance of the remaining two policies (one can see the convergence towards the right of the Figure \ref{fig_p_q1}, and the  percentage improvement is less than 5\%, in both the cases,   when the number  of sources is near 100). In fact, we observe this in many other examples.
   {\it Thus one can use this much simplified blind RR policy when the number of sources is extremely large.} 
   
   \item In all the case studies,  the $\epsilon$-O policy outperforms the other two.  The percentage improvement over the partial information or  PI policy is    up to 80\% (higher when the number of sources is small) and that against blind RR policy is even higher.  
\end{enumerate}

\begin{figure}[h]
   \begin{minipage}{.24\textwidth}
  \includegraphics[trim = {0cm 6cm 1cm 6.5cm}, clip, scale = 0.2]{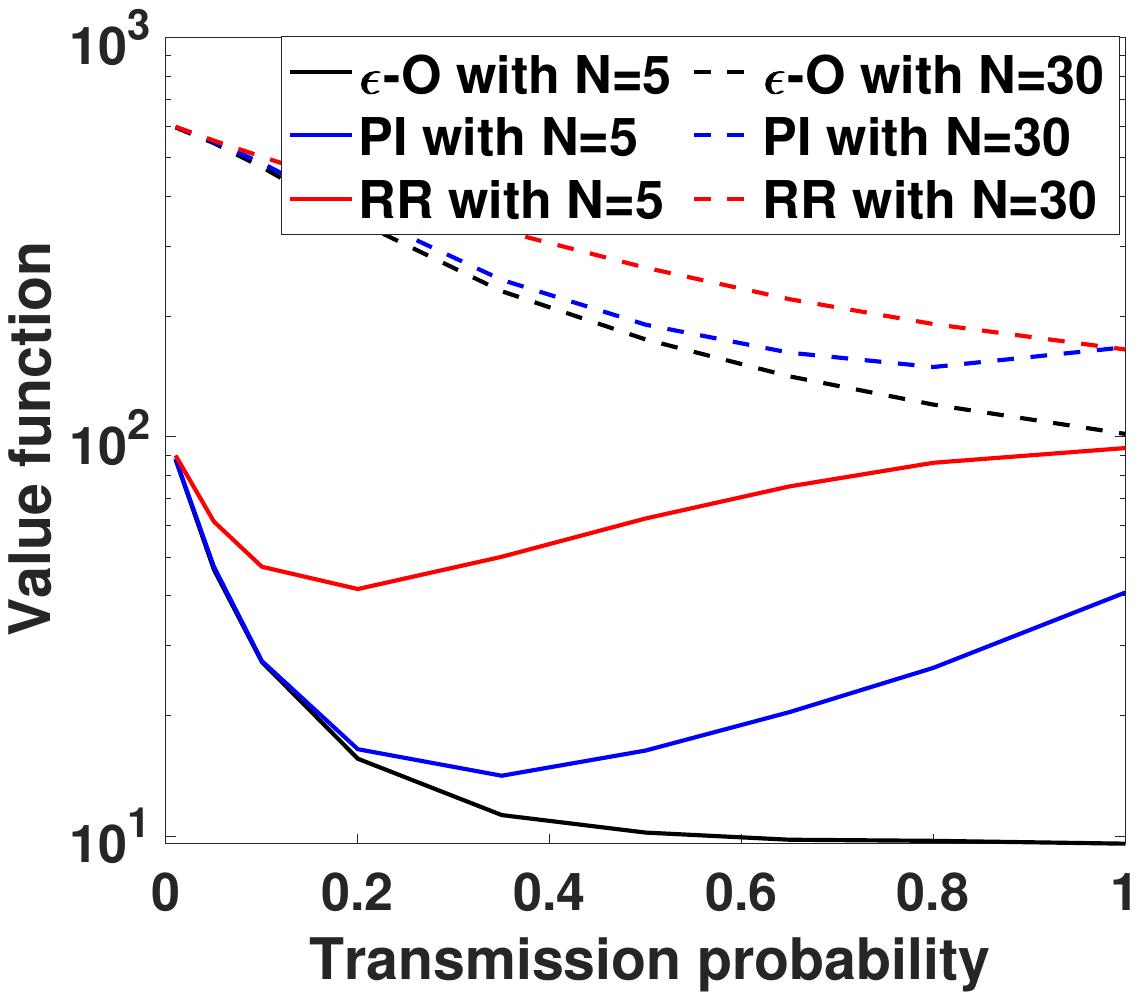}
    \end{minipage}
    \begin{minipage}{.24\textwidth}
   \includegraphics[trim = {0cm 0cm 0cm 0cm}, clip, scale = 0.2]{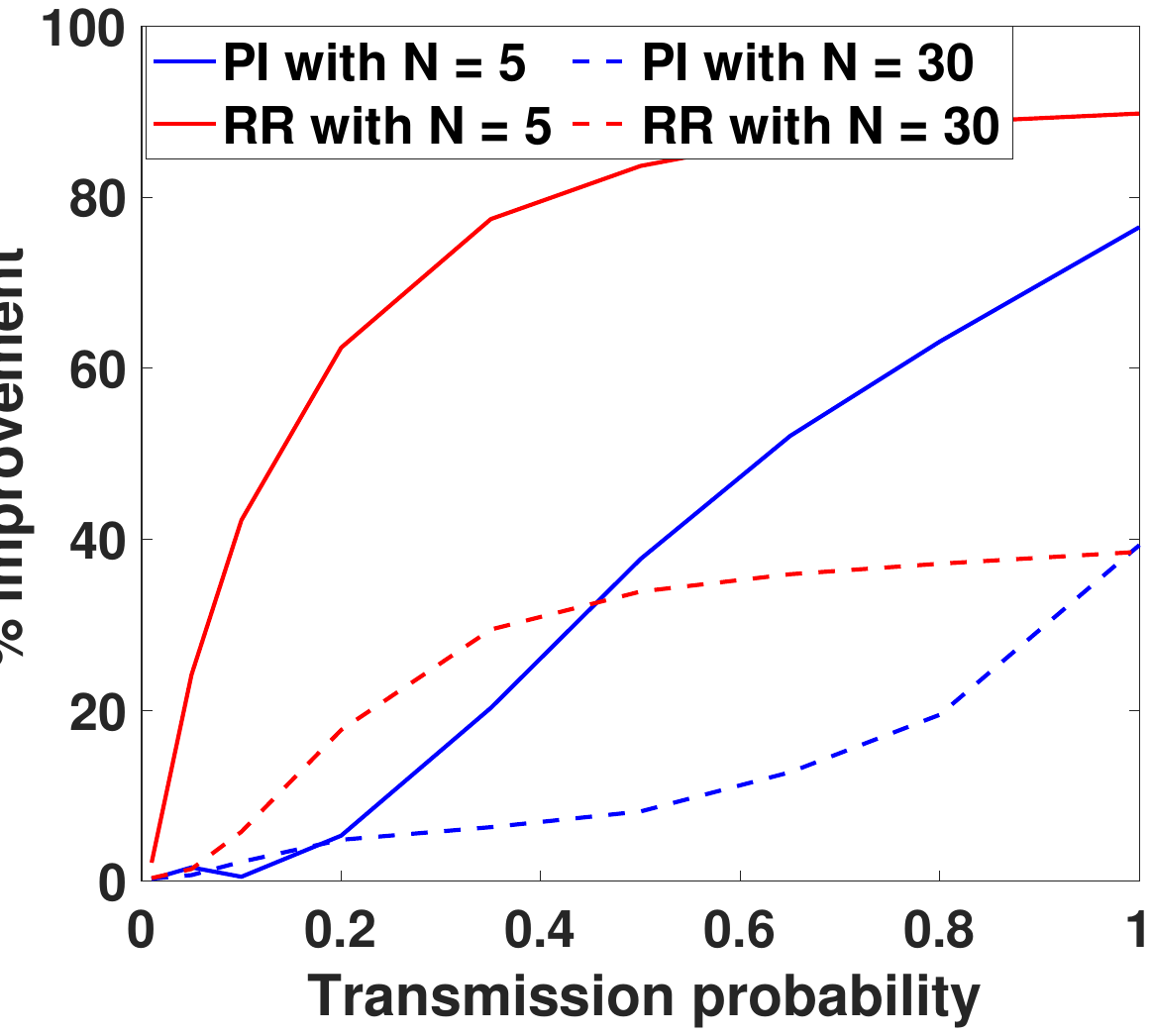} 
    \end{minipage}
    \vspace{-5mm}
    \caption{ {\bf Rare packet arrivals, $q_n = 0.1$ $\forall \ n$:} Value function (on log scale) in left sub-figure  and percentage improvement in the right figure versus transmission probability $p$ with $d=1$   \label{fig_N_q2}}
\end{figure}

\begin{figure}[h]
   \begin{minipage}{.24\textwidth}
    \includegraphics[trim = {0cm 6cm 0cm 7cm}, clip, scale = 0.2]{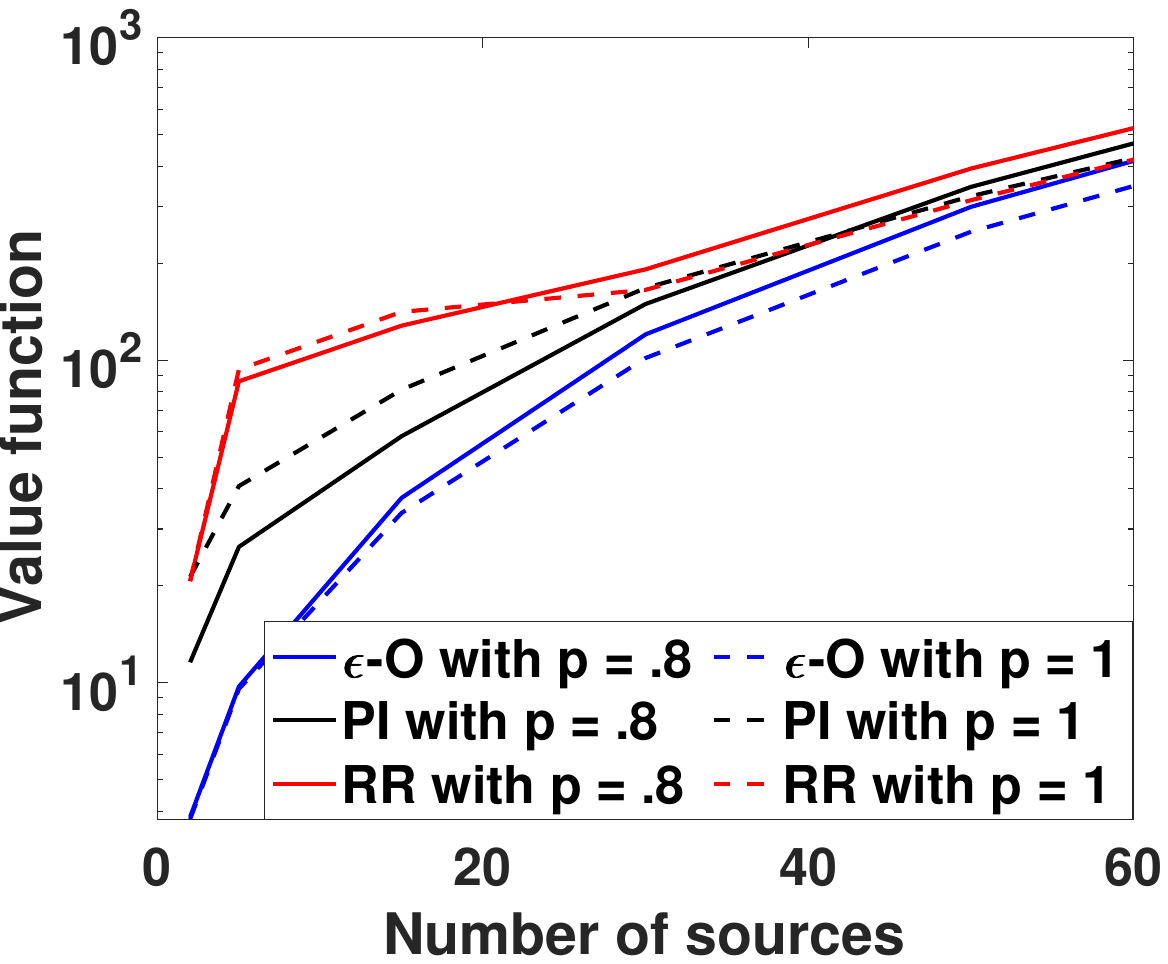}
    \end{minipage}
    \begin{minipage}{.24\textwidth}
    \includegraphics[trim = {0cm 6cm 1cm 7cm}, clip, scale = 0.2]{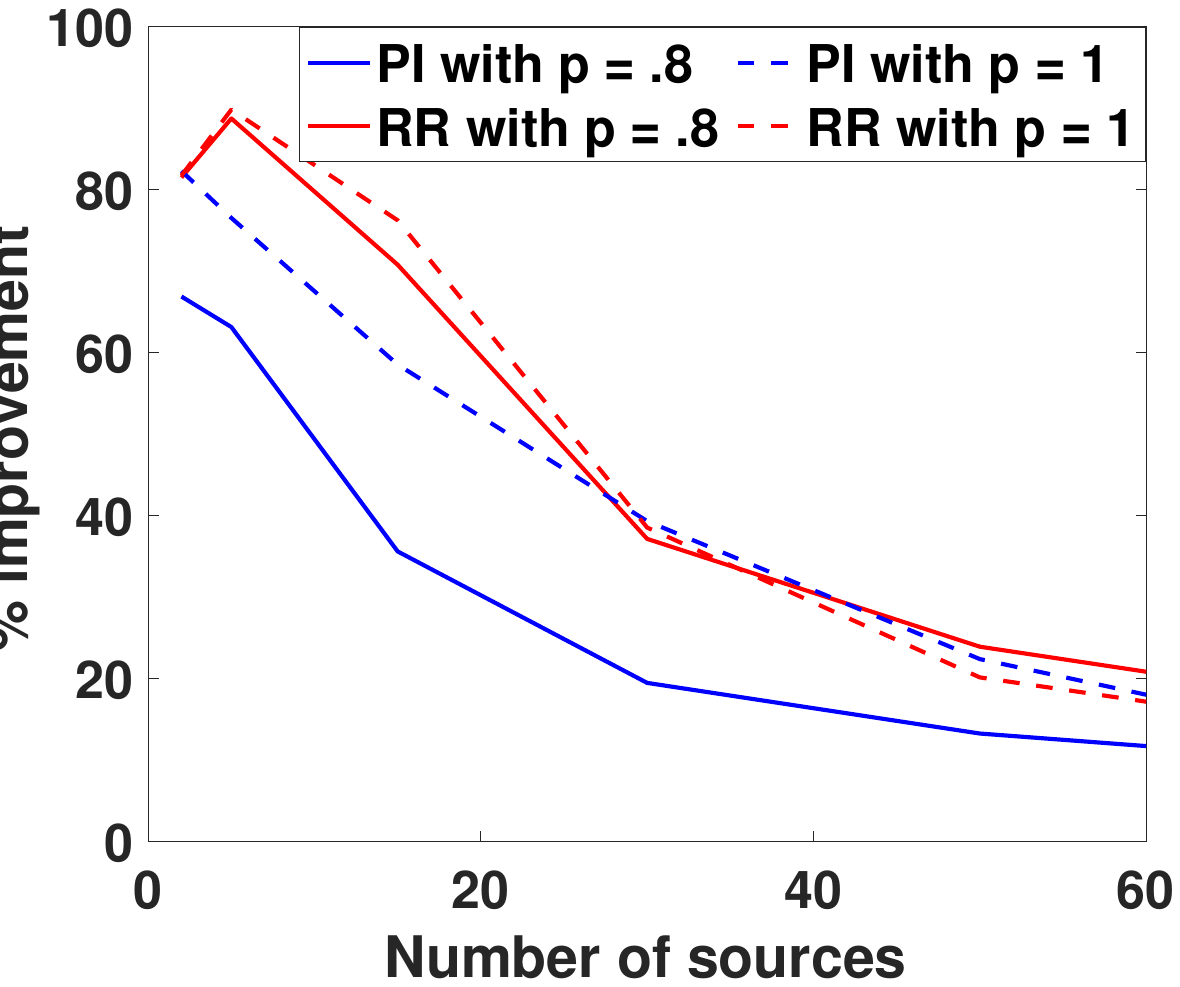}
    \end{minipage}
     \caption{{\bf Rare packet arrivals, $q_n = 0.1$ $\forall \ n$:} Value function (on log scale) in left sub-figure  and percentage improvement in the right figure versus number of sources $N$ with $d=1$   \label{fig_p_q2}}
\end{figure}

Next, we consider an example with even rarer packet arrivals, where  $q_n = .1$ for all $n \in \mathcal{N}$ in Figures \ref{fig_N_q2}-\ref{fig_p_q2}. 
\vspace{-13mm}

In contrast to the previous example, we see even bigger improvements for this rare arrival case study; one can anticipate this as the AoIs at sources also convey significant information in such cases and we managed to derive a simple stationary policy that  uses   both sets of AoI; the increase in the complexity is minimal (one needs to order according the differences in AoIs at the source and the destination as compared to ordering according to AoIs just at the destination),  yet $\epsilon$-O  policy effectively provides significant improvement.

\begin{figure}[h]
   \begin{minipage}{.24\textwidth}
    \includegraphics[trim = {0cm 6cm 0cm 7cm}, clip, scale = 0.2]{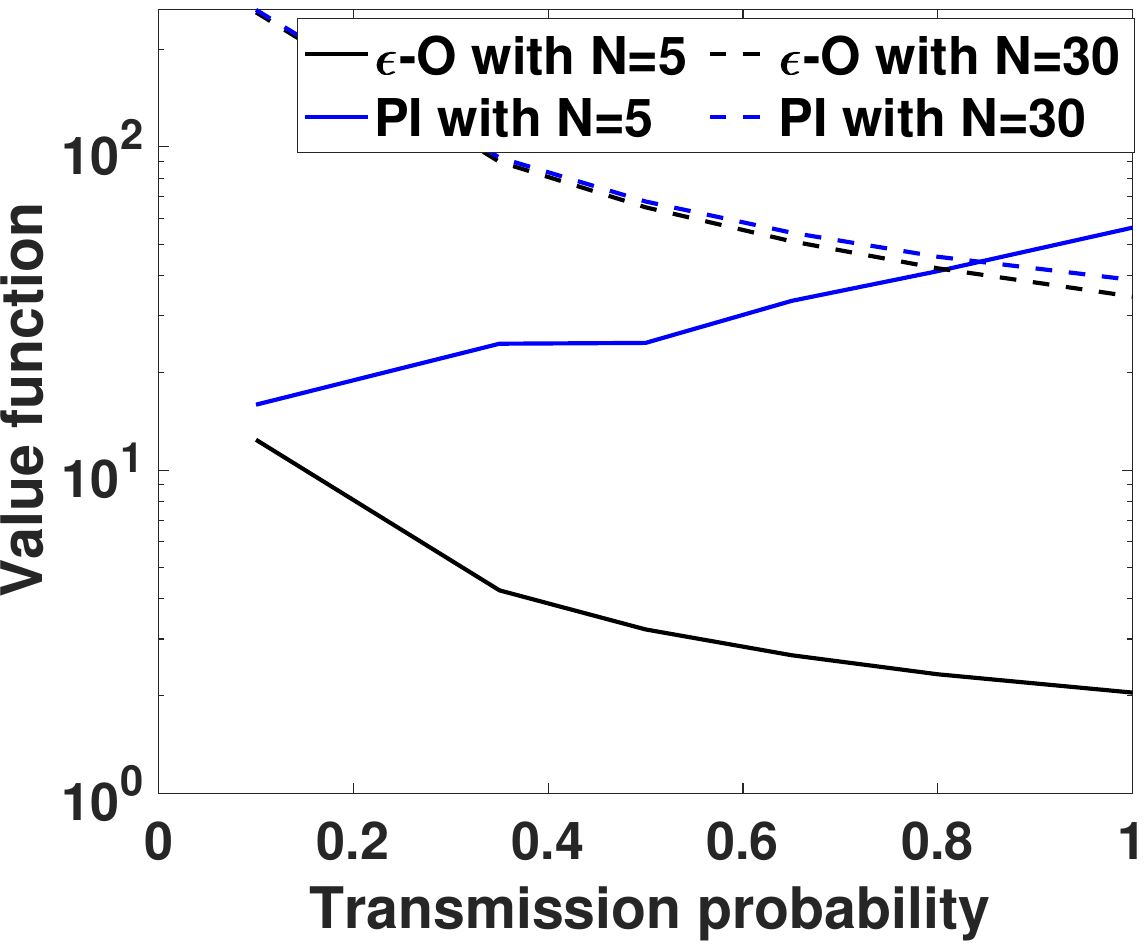}
    \end{minipage}
    \begin{minipage}{.24\textwidth}
    \includegraphics[trim = {0cm 6.5cm 1cm 7.4cm}, clip, scale = 0.2]{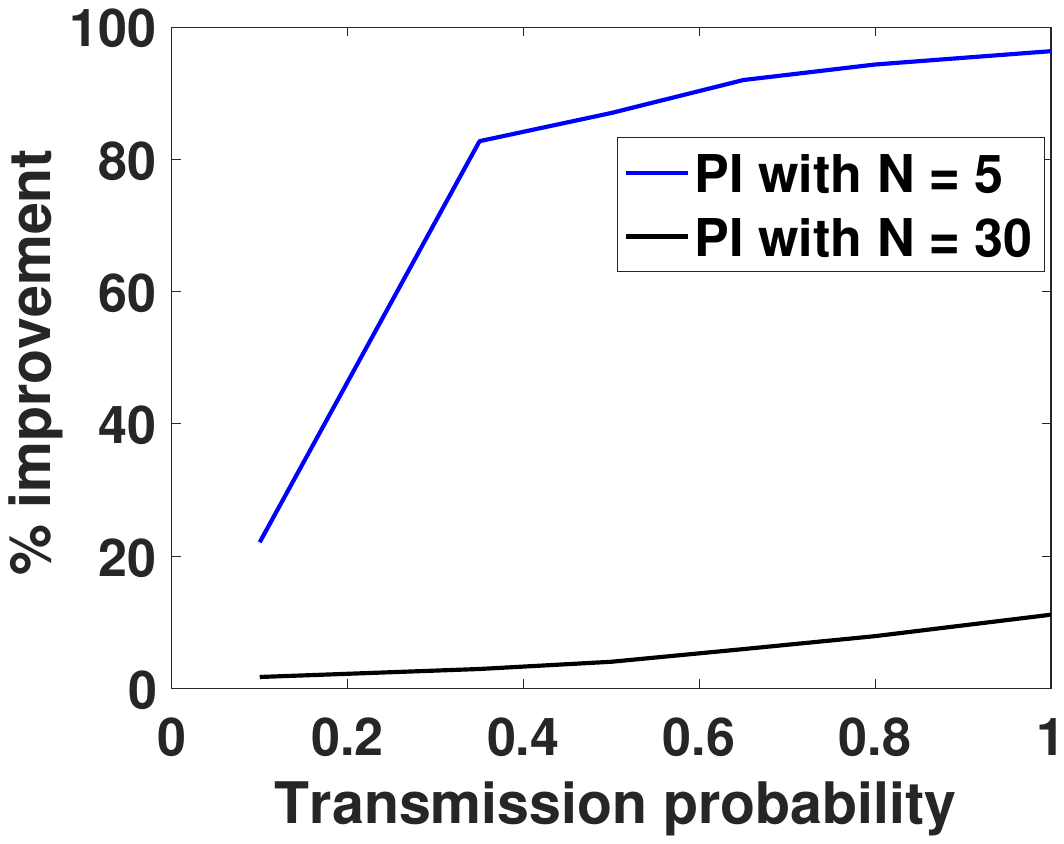}
    \end{minipage}
     \caption{{\bf Packet arrivals, $q_n = 0.5$ $\forall \ n$:} Value function (on log scale) in left sub-figure  and percentage improvement in the right figure versus transmission probability $p$ with $d=3$   \label{fig_N_q1_e}}
\end{figure}

\begin{figure}[h]
   \begin{minipage}{.24\textwidth}
    \includegraphics[trim = {0cm 6.5cm 0cm 7cm}, clip, scale = 0.2]{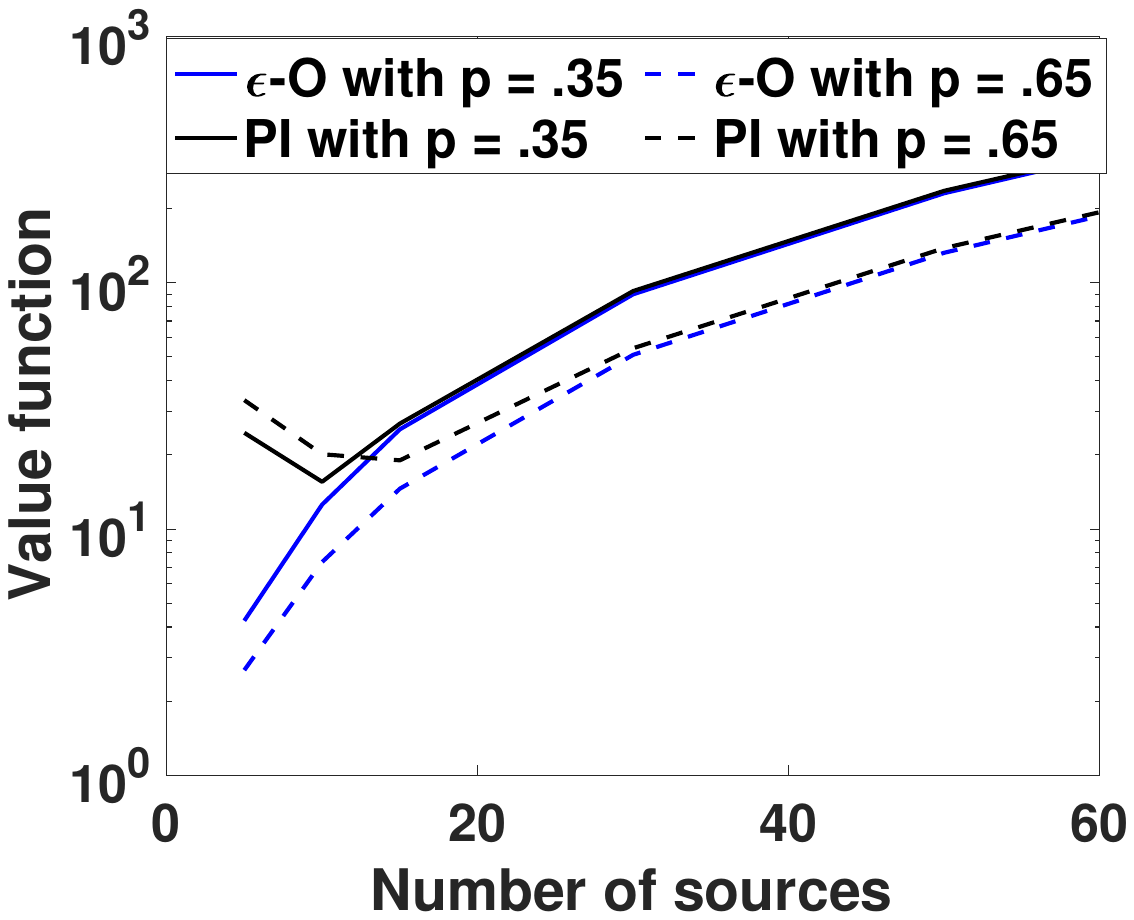}
    \end{minipage}
    \begin{minipage}{.24\textwidth}
    \includegraphics[trim = {0cm 6cm 1cm 7cm}, clip, scale = 0.2]{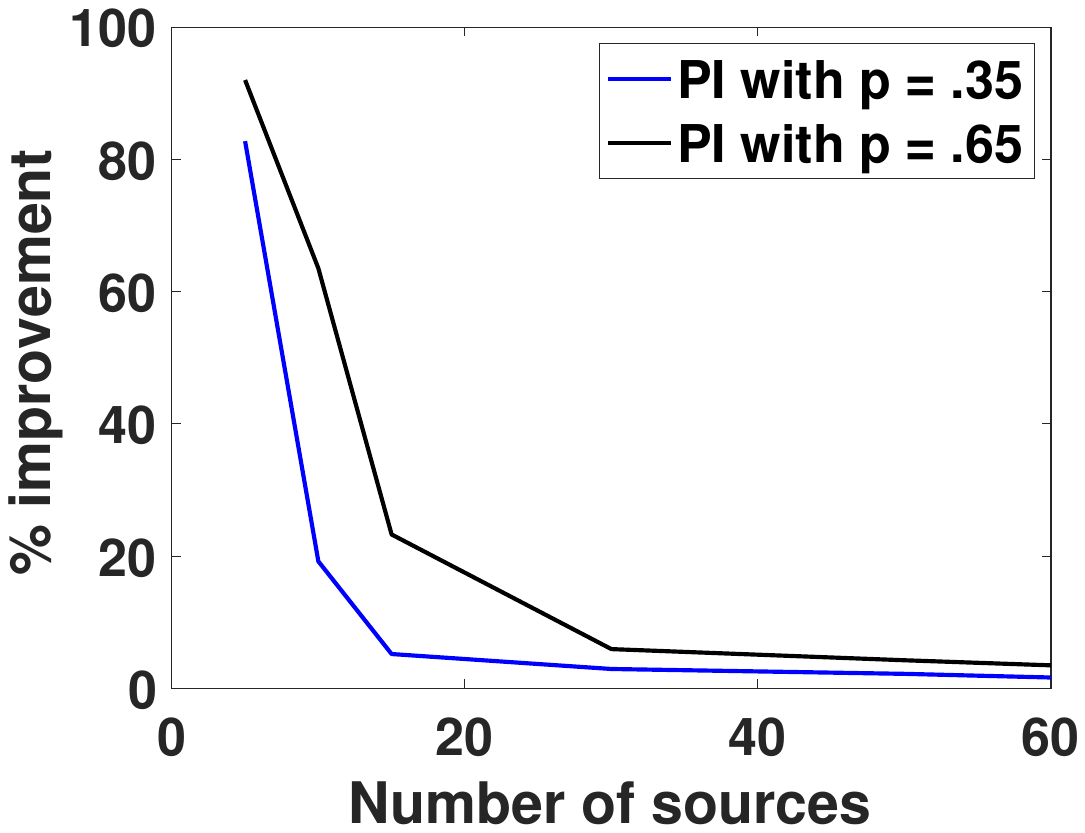}
    \end{minipage}
     \caption{{\bf Packet arrivals, $q_n = 0.5$ $\forall \ n$:} Value function (on log scale) in left sub-figure  and percentage improvement in the right figure versus number of sources $N$ with $d=3$   \label{fig_p_q1_e}}
\end{figure}

\begin{figure}[h]
   \begin{minipage}{.24\textwidth}
    \includegraphics[trim = {0cm 6cm 0cm 7cm}, clip, scale = 0.2]{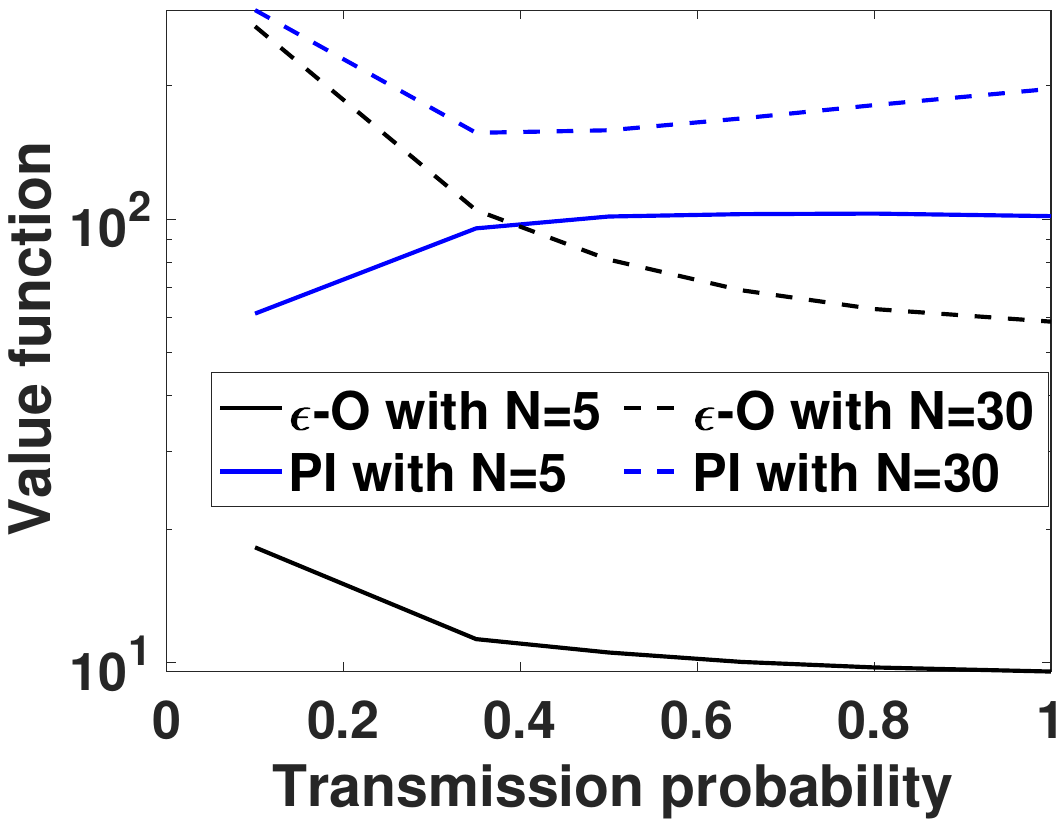}
    \end{minipage}
    \begin{minipage}{.24\textwidth}
    \includegraphics[trim = {0cm 6cm 1cm 7cm}, clip, scale = 0.2]{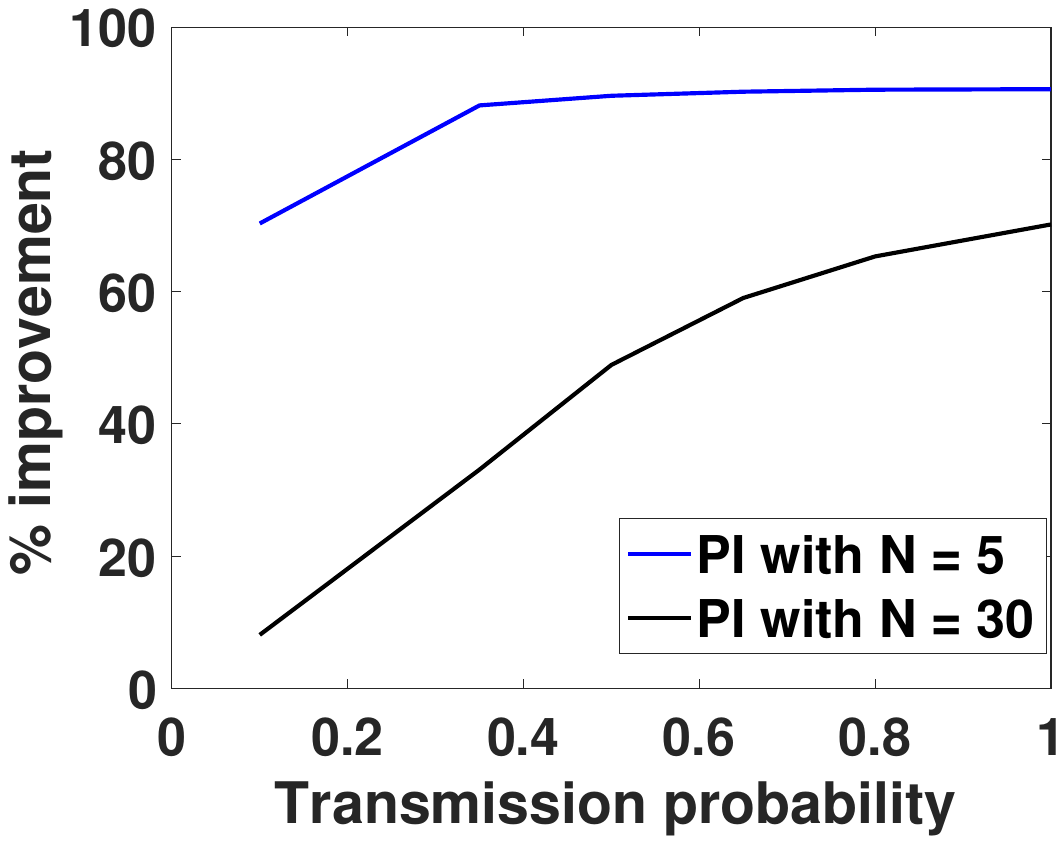}
    \end{minipage}
     \caption{{\bf Rare packet arrivals, $q_n = 0.1$ $\forall \ n$:} Value function (on log scale) in left sub-figure  and percentage improvement in the right figure versus transmission probability $p$ with $d=3$   \label{fig_N_q2_e}}
\end{figure}

\begin{figure}[ht]
   \begin{minipage}{.24\textwidth}
    \includegraphics[trim = {0cm 6cm 0cm 7cm}, clip, scale = 0.2]{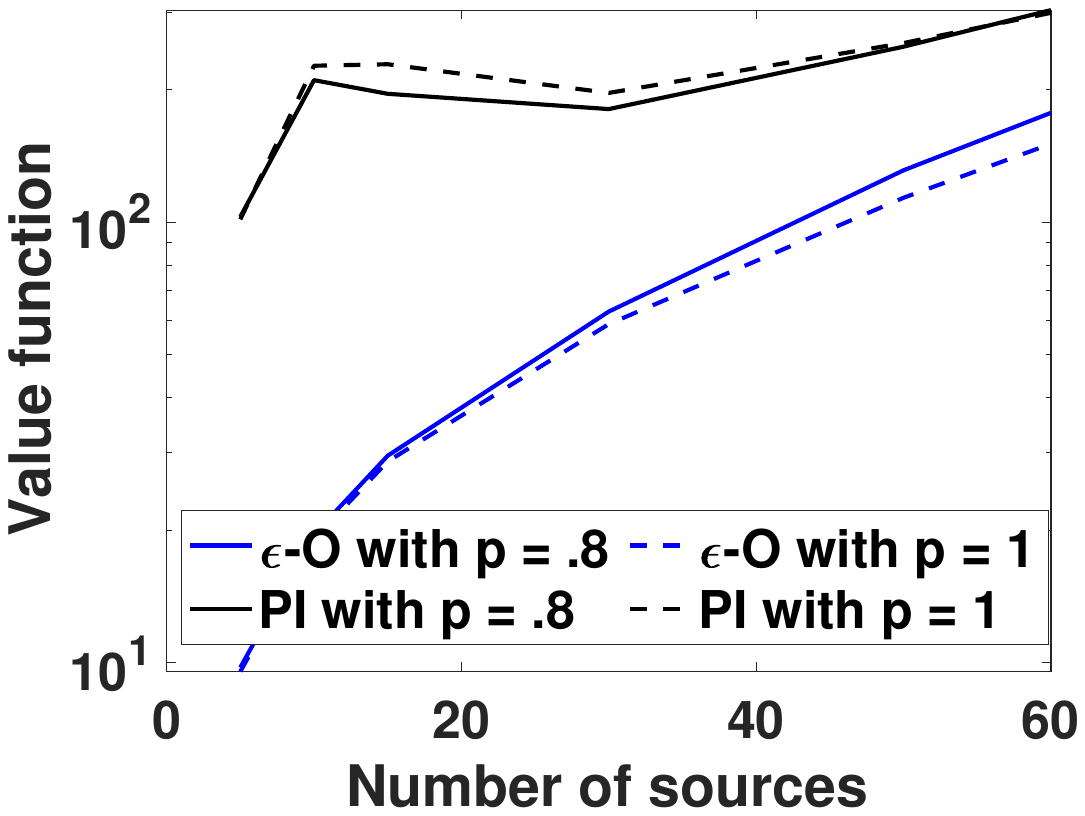}
    \end{minipage}
    \begin{minipage}{.24\textwidth}
    \includegraphics[trim = {0cm 6cm 1cm 7cm}, clip, scale = 0.2]{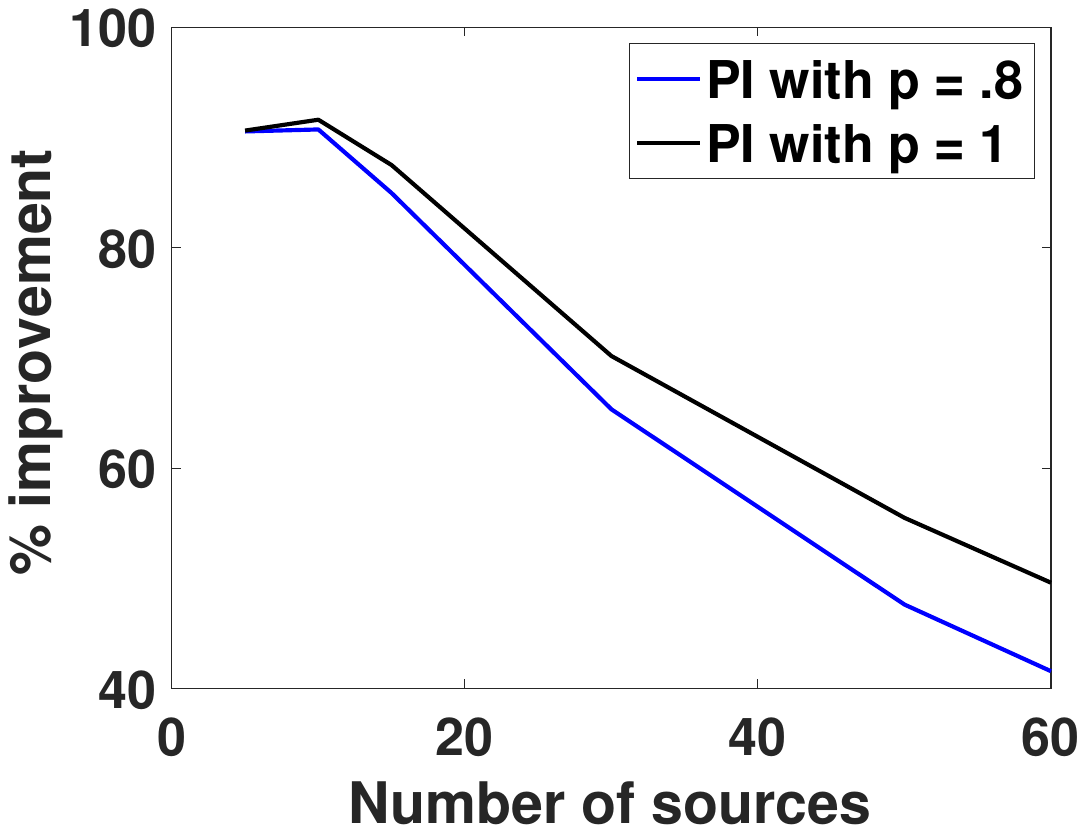}
    \end{minipage}
     \caption{{\bf Rare packet arrivals, $q_n = 0.1$ $\forall \ n$:} Value function (on log scale) in left sub-figure  and percentage improvement in the right figure versus number of sources $N$ with $d=3$   \label{fig_p_q2_e}}
\end{figure}

Similar trends are observed for other case studies with multiple channels (with $d=3$) in  Figures \ref{fig_N_q1_e}-\ref{fig_p_q2_e};   now the improvements are even higher, for example in the right sub-figure of Figure \ref{fig_N_q1_e}  when $p$ is close to $1$ and the number of sources   $N=30$, the improvement is up to $90$\%, even for the case with reasonable packet arrival rates, i.e., with $q_n = .5$.

\section{Conclusion}
We consider a system with multiple sources trying to transmit information packets to a common destination via multiple orthogonal channels. Each source has its own buffer with single storage capacity. The packets arrive to each of the sources according to a Bernoulli process while the transfer times are geometric. We formulate it as a finite horizon Markov Decision Process (MDP) and derive a near optimal policy. Interestingly, the derived policy (defined only in terms of the differences in AoIs at the sources and the destination) is stationary  and further does not depend on the packet arrival rates to any of the sources; this additional computational advantage is mainly due to the fact that the policy is near optimal. We demonstrate the superiority of the proposed policy by comparing it with the adaptations of the existing policies in literature for our case with unreliable packet generation, through numerical experiments.
There are several future directions here, for example, to study the source selection policy for multiple sources transmitting packets to multiple destinations via multiple channels.

\bibliographystyle{unsrt}
\bibliography{ACC/references}

\section*{Appendix}

\noindent
\textbf{Proof of Theorem \ref{thm_eps_optimal}:} 
%
We directly compute the difference between the objective function under the optimal policy (i.e, value function) and that under the policy \eqref{Eqn_eps_opt_policy}. We derive it by solving the following DP equations using backward recursion. 
\begin{eqnarray}
\begin{aligned}
V_{T}^*(x) & =& c(x) \text{ for all } x  \text{ and for any } t < T, \ \forall \ x, \\
  V_{t}^*(x) & = & \min_{{\bf a} \in \mathcal{A}_x} \left[ c(x,{\bf a}) + \mathbb{E} \big[ V_{t+1}^*(X')|x,{\bf a}\right].
  \end{aligned}
  \label{Eqn_DP}
\end{eqnarray}
Recall now ${\bf a} = (a_1, a_2 \cdots, a_{N_x^d})$   is the subset of sources that attempted transition at state $x$ in the previous slot.

\noindent
\textbf{At terminal epoch, i.e., at $t = T$: }There are no further transmission attempts and hence from \eqref{Eqn_cost} and \eqref{Eqn_DP},
\begin{eqnarray}
\label{Eqn_VT}
    V_T^*(x) = c(x) = \sum_{n \in {\cal N}} h_n \text{ for all } x.  
\end{eqnarray}

\noindent
\textbf{At decision epoch  $t = T-1$: }Towards computing value function, one needs to derive Q-functions corresponding to each possible action, and then
\begin{eqnarray}
    V_{T-1}^*(x) = 
    \min_{{\bf a} \in \mathcal{A}_x}  Q_{T-1}(x,{\bf a}). 
    \label{Eqn_V_fromQ}
\end{eqnarray}
Consider any $x $ and any ${\bf a} \in \mathcal{A}_x$. We begin with computing $Q(x,{\bf a})$. Towards that,
let $H_n'$ be the age of the packet corresponding to source $n$ at destination in time slot $T$ when sources in subset ${\bf a}$ are chosen for transfer -- observe that $H_n' = h_n+1$ for all $n \notin {\bf a}$ and otherwise $H_n' = g_n+1$ or $h_n+1$ depending upon the success of the transfer of packet from source $n$. Thus from \eqref{Eqn_DP}-\eqref{Eqn_V_fromQ},

\vspace{-4mm}
{\small \begin{eqnarray*}
Q_{T-1}(x,{\bf a}) \ = \ 
\sum_{n \in \mathcal{N}} h_n +  \mathbb{E} \left[ \sum_{n \in \mathcal{N}} H_n'|x,{\bf a}\right], \hspace{-50mm} \\
    & = & \hspace{-2mm} 2\sum_{n \in \mathcal{N}}h_n + N+p l(x,\a) \text{ with},  \\ l(x,\a)  &:=&\sum_{i=1}^{N_x^d}(g_{a_i}-h_{a_i}),
\end{eqnarray*}}%
where the last equality follows from Lemma \ref{lem_expect}.
%
%
\TR{For any $x \in {\cal X}_\psi$,   no source have any packet to transmit and thus, $H_n' = h_n+1$ for all $n$. Again using \eqref{Eqn_DP}-\eqref{Eqn_V_fromQ} we obtain,

\vspace{-6mm}
{\small \begin{eqnarray*}
 \hspace{-5mm}   Q_{T-1}(x,a_\psi) 
     & \hspace{-2mm} = \hspace{-2mm}& \sum_{n \in \mathcal{N}} h_n +  \mathbb{E} \left[ \sum_{n \in \mathcal{N}} H_n'|x,a_\psi\right] \\
\hspace{-5mm}    & \hspace{-2mm}=\hspace{-2mm} & \sum_{n \in \mathcal{N}}h_n +  \sum_{n \in \mathcal{N}}(h_n+1) 
= 2\sum_{n \in \mathcal{N}}h_n + N.
\end{eqnarray*}}}{} 
Hence, using \eqref{Eqn_eps_opt_policy} with
$
l(x) = l^\Delta(x) := \min_{{\bf a} \in {\cal A}_x} l(x,\a),
$%
we have for all $x$
 \begin{eqnarray}
    V^*_{T-1}(x) \ = \ 
     2\sum_{n \in \mathcal{N}}h_n + N+p l^\Delta(x) \ = \ V^\Delta_{T-1}(x).
     \label{Eqn_V_T1}
\end{eqnarray}
Observe that  the last equality follows as the decisions while defining $V^*_{T-1}$ are the same as in that in $\pi^\Delta$ of \eqref{Eqn_eps_opt_policy}.

\noindent
\textbf{At decision epoch $t=T-2$: }
We again compute the Q-functions  using \eqref{Eqn_V_T1} and with $\{H_n'\},\{G_n'\},X'$ now representing the quantities at $T-1$. For any $x$ and ${\bf a} \in \mathcal{A}_x$,  

\vspace{-2.5mm}
{\small \begin{eqnarray*}
Q_{T-2}(x,{\bf a}) \hspace{8mm} \\ 
      & \hspace{-30mm} =& \hspace{-15mm} \sum_{n \in \mathcal{N}}h_n + \mathbb{E} \bigg[ \bigg( 2\sum_{n \in \mathcal{N}}H_n' + N \left. +p l^\Delta(X') \bigg) \right|  x,{\bf a} \bigg], \\
       & \hspace{-30mm} = &  \hspace{-15mm} 3\sum_{n \in \mathcal{N}}h_n + 3N+2pl(x,\a)   + p\mathbb{E}[l^\Delta(X')|x,{\bf a}], \\ 
     &\hspace{-30mm}  =& \hspace{-15mm}3\sum_{n \in \mathcal{N}}h_n + 3N+2  p l(x,\a) + p(1-p_x^d){\cal U}(x)\\
       & \hspace{-30mm} & 
      \hspace{-15mm}  + p\cdot p_d {\cal V}(x,{\bf a}),
    \end{eqnarray*}}

  \noindent
 \noindent 
 where the last equality follows from simple algebra and Lemma \ref{lem_F_G} with $\mathcal{U}(x)$ and $\mathcal{V}(x,{\bf a})$ as defined in Lemma \ref{lem_F_G}. \TR{Similarly for any $x \in {\cal X}_\psi$, using Lemma \ref{lem_F_G}.(ii) we have

 \begin{eqnarray*}
Q_{T-2}(x,a_\psi) 
  & = & 3\sum_{n \in \mathcal{N}}h_n + 3N   + p\mathcal{U}(x).   
  \end{eqnarray*}}{}
 \noindent
 Thus for any $x$, 
 
 \vspace{-2mm}
 {\small \begin{eqnarray*}
  V^*_{T-2}(x) 
   & = &  3\sum_{n \in \mathcal{N}}h_n + 3N + p(1-p_x^d) {\cal U}(x)\\
  && \hspace{-10mm}  + p\min_{{\bf a} \in \mathcal{A}_x} \bigg[ 2 l(x,{\bf a})   + p_d \mathcal{V}(x,{\bf a})\bigg]. 
  \end{eqnarray*}}

  \noindent
Similarly, under $\pi^\Delta$ of \eqref{Eqn_eps_opt_policy}, with ${\bf a}^\Delta =d^\Delta(x) $, 
 $
  V^\Delta_{T-2}(x) 
  =  Q_{T-2}(x,\a^\Delta). 
 $
Now with $l^*(x) := l(x,\a^*)$ and ${\bf a}^* := \arg \min_{{\bf a} \in \mathcal{A}_x} \left[ 2l(x,\a)   + p _d\mathcal{V}(x,{\bf a})\right] $,

\vspace{-2mm}
 {\small  \begin{eqnarray*}
 \left | V^{\Delta}_{T-2}(x) - V^*_{T-2}(x) \right | 
& = & V^{\Delta}_{T-2}(x) - V^*_{T-2}(x), \\
& \hspace{-17mm} = & \hspace{-10mm} 2p\big[ l^\Delta(x)-l^*(x) \big] \\
&& \hspace{-10mm} + p \cdot p_d \big(\mathcal{V}(x,{{\bf a}^\Delta})-\mathcal{V}(x,{\bf a}^*) \big), \\
& \hspace{-17mm} \le & \hspace{-10mm} p\cdot p_d  \big(\mathcal{V}(x,{{\bf a}^\Delta})-\mathcal{V}(x,{\bf a}^*) \big), \\
& \hspace{-17mm} \le &  \hspace{-10mm} p\cdot p_d \left( 2 d || x||_\infty \right),
\end{eqnarray*}}%
where the last inequality again follows from Lemma \ref{lem_F_G}.  Hence at $t= T-2$, 
  $V^*_{t}(x) =  V^\Delta_{t}(x)-p\cdot p_d  \fun_{t}(x)$
where the difference function is bounded as
$0 \le \fun_{t}(x) \le 2d|| x||_\infty $.

\noindent
Next, assume that the value function and the objective under policy $\pi^\Delta$ has the following form for any $k\ge2$,

\vspace{-2mm}
{\small \begin{eqnarray}
    \label{eqn_val_func}
    V^*_{T-k}(x) & \hspace{-2mm} = \hspace{-2mm}& V^\Delta_{T-k}(x) -p\cdot p_d \fun_{T-k}(x), \\
    \label{eqn_val_eps}
    V^\Delta_{T-k}(x)& \hspace{-2mm}= \hspace{-2mm}& (k+1) \sum_{n \in \mathcal{N}} h_n + K(k)+ p{\tilde V}_{T-k}(x) \text{ with } \\
    \label{eqn_constant}
    K(k) &\hspace{-2mm} =\hspace{-2mm} & kN + K(k-1), \\
    \label{eqn_common}
  {\tilde V}_{T-k}(x) &\hspace{-2mm} = & {\cal F}_{T-k}(x)  + kl^\Delta(x)
\end{eqnarray}}%
for some appropriate functions, ${\cal F}_{T-k}$ and ${\cal Z}_{T-k}$  which can be upper bounded as below,

\vspace{-4mm}
{\small \begin{eqnarray}
\label{eqn_bound}
{\cal F}_{T-k}(x)  &\hspace{-2mm} \le \hspace{-2mm} & C_1(k)||x||_\infty  + C_2(k), \\
\label{eqn_Z_bd}
{\cal Z}_{T-k}(x)  & \hspace{-2mm} \le \hspace{-2mm} & D_1(k) ||x||_\infty + D_2(k),  \mbox{ with, } 
 \\
\label{eqn_C1}
C_1(k) & \hspace{-2mm} = \hspace{-2mm}  & (1+p_d) C_1(k-1)+2(k-1)d,\\\
\label{eqn_C2}
C_2(k) & \hspace{-2mm} = \hspace{-2mm} & (1+p_d) \big[C_1(k-1)+C_2(k-1)\big],\\
\label{eqn_D1}
\hspace{-5mm} D_1(k) & \hspace{-3mm} = \hspace{-3mm} & (1+p_d)D_1(k-1)+2C_1(k-1)+2(k-1)d, \\
\label{eqn_D2}
D_2(k) & \hspace{-2mm} = \hspace{-2mm} & (1+p_d) \big[ D_1(k-1)+D_2(k-1)\big] \nonumber \\
&\hspace{-2mm} \hspace{-2mm} & + 2 \big[ C_1(k-1)+C_2(k-1)\big]. 
\end{eqnarray}  }
Observe that for $k=2$, \eqref{eqn_val_func}-\eqref{eqn_D2} are satisfied with,
\begin{eqnarray*}
    {\cal F}_{T-2}(x) & \hspace{-3mm}= \hspace{-3mm}&(1-p_x^d) {\cal U}(x)+\cdot p_d{\cal V}(x,{\bf a}^\Delta), \ K(1)  \ = \ N,\\
    {\cal Z}_{T-2}(x) & \hspace{-3mm} =  \hspace{-3mm}&  \frac{2}{p_d} \big[l^\Delta(x)-l^*(x) \big]    +\mathcal{V}(x,{\bf a}^\Delta)- \mathcal{V}(x,{\bf a}^*).  
\end{eqnarray*}
and  ${\cal F}_{T-2}(x) \le (1+p_d)d||x||_\infty$ (with  $C_1(2) = (1+p_d)d$ and $C_2(2) = 0$) by Lemma \ref{lem_F_G}, which satisfies \eqref{eqn_bound}; it is already proved that the bound on ${\cal Z}_{T-2}(x)$ (with $D_1(2) = 2d$  and $D_2(2) = 0$) satisfies \eqref{eqn_Z_bd}.
By backward mathematical induction, it suffices to show that  the value function and the objective have the above form at
$t = T-(k+1)$.
For any $x$ and ${\bf a} \in \mathcal{A}_x$ using Lemma \ref{lem_expect}, 

\vspace{-4mm}
{\small \begin{eqnarray}
\label{eqn_Q}
Q_{T-(k+1)}(x,{\bf a}) 
\hspace{-12mm} \nonumber \\
     & = & \hspace{-5mm} \sum_{n \in \mathcal{N}}h_n + \mathbb{E} \bigg[ (k+1)\sum_{n \in \mathcal{N}}H_n' + K(k) + p{\tilde V}_{T-k}(X') \nonumber \\
     &&  -p \cdot p_d \fun_{T-k}(X') \left | x,{\bf a} \bigg] \right., \nonumber\\
    & = & (k+2)\sum_{n \in \mathcal{N}}h_n + K(k+1) +  (k+1)pl(x,\a) \nonumber \\
    && + p\mathbb{E} \bigg[ {\tilde V}_{T-k}(X')- p_d \fun_{T-k}(X') \left | x,{\bf a} \bigg] \right.,
\end{eqnarray}}%
with $K(k+1) = (k+1)N+K(k)$ which satisfies same equation as \eqref{eqn_constant}, but at $k+1$. Further using \eqref{eqn_common} and  Lemma \ref{lem_F_G} (with ${\cal U}(x)$ and ${\cal V}(x,{\bf a})$ defined there and bounded by $d||x||_\infty$) the last but one term of \eqref{eqn_Q}, 
%

\vspace{-2mm}
{\small \begin{align*}
\mathbb{E} \left[ {\tilde V}_{T-k}(X') | x,{\bf a} \right] \hspace{-17mm} \\
   & =  \mathbb{E} [{\cal F}_{T-k}(X')|x,{\bf a}]   +k \big[(1-p_x^d) {\cal U}(x) + p_d{\cal V}(x,{\bf a})\big]. 
\end{align*}}
 Substituting the above in \eqref{eqn_Q}, we obtain,

\vspace{-4mm}
 {\small
  \begin{eqnarray}
 Q_{T-(k+1)}(x,{\bf a}) 
     =     (k+2)\sum_{n \in \mathcal{N}}h_n + K(k+1) +  (k+1)pl(x,\a) \hspace{-54mm}\nonumber \\
    && \hspace{-30mm} + p\mathbb{E} [{\cal F}_{T-k}(X')|x,{\bf a}]   +kp \big[(1-p_x^d) {\cal U}(x) + p_d{\cal V}(x,{\bf a})\big] \nonumber \\
    \label{eqn_q_k1}
    && \hspace{-30mm}-p \cdot p_d  \mathbb{E} \big[  \fun_{T-k}(X') \left | x,{\bf a} \big] \right..
 \end{eqnarray}}
\TR{For any $x \in {\cal X}_\psi$, proceeding as before and using Lemma \ref{lem_F_G}.(ii) we obtain,

\vspace{-4mm}
{\small\begin{eqnarray*}
Q_{T-(k+1)}(x,a_\psi) \hspace{-20mm} \\
     & = & (k+2)\sum_{n \in \mathcal{N}}h_n + K(k+1)  +p\mathbb{E} [{\cal F}_{T-k}(X')|x,a_\psi] \\
     && +kp{\cal U}(x)-p^2 \mathbb{E} \left[ \fun_{T-k}(X')|x,a_\psi \right]
\end{eqnarray*}}}{}
\noindent
Conditioning on $\mathbb{S}$ (flag indicating at least one successful packet transfer) and following steps exactly as in Lemma \ref{lem_F_G}, we get functions (which depend on time) ${\cal G}_{T-k}(x)$, ${\cal G}'_{T-k}(x)$, ${\cal H}_{T-k}(x,{\bf a})$, and ${\cal H}'_{T-k}(x,{\bf a})$, one can show that, 
\begin{align}
    \mathbb{E}[\mathcal{F}_{T-k}(X')|x,{\bf a}] & = (1-p_x^d)\mathcal{G}_{T-k}(x)+ p_d\mathcal{H}_{T-k}(x,{\bf a}), \nonumber \\
    \mathbb{E}[\mathcal{Z}_{T-k}(X')|x,{\bf a}] & = (1-p_x^d)\mathcal{G}'_{T-k}(x)  + p_d\mathcal{H}'_{T-k}(x,{\bf a}). \nonumber 
\end{align}
Further observe $||X'||_\infty \le ||x||_\infty + 1$ a.s. irrespective of ${\bf a}$ and hence, ${\cal F}_{T-k}(X') \le  C_1(k)(||x||_\infty+1)  + C_2(k)$ using \eqref{eqn_bound}. Conditioning on ${\mathbb S}=0$, we have,
\begin{equation*}
    \mathbb{E} [ {\cal F}_{T-k}(X')|x,{\bf a},{\mathbb S}=0] = {\cal G}_{T-k}(x).
\end{equation*}
Thus we have $|{\cal G}_{T-k}(x)| \le  C_1(k)(||x||_\infty+1)  + C_2(k)$, from above and using \eqref{eqn_bound}. Similar argument follows and the upper bound for ${\cal H}_{T-k}(x,\a)$ matches with that on ${\cal G}_{T-k}(x)$; further ${\cal G}'_{T-k}(x)$ and ${\cal H}'_{T-k}(x,{\bf a})$ can be upper bounded with $D_1(k)(||x||_\infty+1)  + D_2(k)$.
Hence from \eqref{eqn_q_k1},

\vspace{-4mm}
{\small\begin{align}
V^*_{T-(k+1)}(x)   
    & =  (k+2)\sum_{n \in \mathcal{N}}h_n + K(k+1)   +p(1-p_x^d){\cal G}_{T-k}(x)\nonumber \\ & 
     + kp(1-p_x^d){\cal U}(x) -p\cdot p_d(1-p_x^d) {\cal G}'_{T-k}(x) \nonumber \\
    &  + p \min_{{\bf a} \in {\cal A}_x} \bigg[ (k+1) l(x,\a)  +p_d {\cal H}_{T-k}(x,{\bf a}) \nonumber \\
     \label{eqn_val_k}
    & -(p_d)^2 {\cal H}'_{T-k}(x,{\bf a})+kp_d {\cal V}(x,{\bf a})\bigg].
\end{align}}
\noindent
In similar lines using \eqref{eqn_val_eps}-\eqref{eqn_common}, 

\vspace{-4mm}
{\small\begin{align}
V^\Delta_{T-(k+1)}(x)   
     & =  (k+2)\sum_{n \in \mathcal{N}}h_n + K(k+1)   +p(1-p_x^d){\cal G}_{T-k}(x)\nonumber \\ & 
     + kp(1-p_x^d){\cal U}(x) 
      + (k+1)pl^\Delta(x) \nonumber \\
      \label{eqn_obj_k}
      & +  p\cdot p_d \bigg[ {\cal H}_{T-k}(x,{\bf a}^\Delta) 
 +k {\cal V}(x,{\bf a}^\Delta)\bigg].
\end{align}}%
Now, the difference between the objective under policy of \eqref{Eqn_eps_opt_policy} and value function is bounded as below,

\vspace{-6mm}
{\small \begin{eqnarray}
p \cdot p_d \fun_{T-(k+1)} (x) := 
 V^\Delta_{T-(k+1)}(x)    - V^*_{T-(k+1)}(x), \hspace{-55mm} \nonumber \\
 & \hspace{-30mm}= & \hspace{-15mm} p \cdot p_d(1-p_x^d){\cal G}'_{T-k}(x) \nonumber   + (k+1)p\big[  l^\Delta(x) - l^*(x)\big] \nonumber \\
    & \hspace{-30mm} & \hspace{-15mm} + p \cdot p_d \bigg[   {\cal H}_{T-k}(x,{\bf a}^\Delta)- {\cal H}_{T-k}(x,{\bf a}^*) \nonumber \\
    & \hspace{-30mm}& \hspace{-15mm} +k\big[ {\cal V}(x,{\bf a}^\Delta)-{\cal V}(x,{\bf a}^*)\big]+p_d{\cal H}'_{T-k}(x,{\bf a}^*) \bigg], \nonumber \\
    & \hspace{-30mm} \le & \hspace{-15mm} p \cdot p_d(1-p_x^d){\cal G}'_{T-k}(x) \nonumber  \\
    & \hspace{-30mm} & \hspace{-15mm} + p \cdot p_d \bigg[   {\cal H}_{T-k}(x,{\bf a}^\Delta)- {\cal H}_{T-k}(x,{\bf a}^*) \nonumber \\
    & \hspace{-30mm} & \hspace{-15mm} +k\big[ {\cal V}(x,{\bf a}^\Delta)-{\cal V}(x,{\bf a}^*)\big]+p_d{\cal H}'_{T-k}(x,{\bf a}^*) \bigg], \nonumber \\
    & \hspace{-30mm} \le & \hspace{-15mm} p \cdot p_d \bigg[(1+p_d)\big[D_1(k)(||x||_\infty+1)+D_2(k)\big] \nonumber \\
    & \hspace{-30mm} & \hspace{-15mm} + 2\big[C_1(k)(||x||_\infty+1)+C_2(k)\big]+2kd||x||_\infty\bigg], \nonumber \\
    & \hspace{-30mm} = & \hspace{-15mm} p \cdot p_d \bigg[  \big[(1+p_d)D_1(k)+2C_1(k) +2kd \big] ||x||_\infty  \nonumber \\
    \label{eqn_Z_diff}
    & \hspace{-30mm} & 
   \hspace{-15mm} + (1+p_d) \big[ D_1(k) + D_2(k) \big] +2 \big[ C_1(k) + C_2(k) \big] \bigg],
\end{eqnarray}}%
which thus satisfies  \eqref{eqn_val_func} and \eqref{eqn_Z_bd} using the recursive constants defined as in \eqref{eqn_C1}-\eqref{eqn_D2}. 
 Comparing \eqref{eqn_obj_k} with \eqref{eqn_val_eps}-\eqref{eqn_common}, the function ${\cal F}_{T-(k+1)}$  can be identified and bounded as below,
 
\vspace{-4mm}
{\small \begin{eqnarray*}
   {\cal F}_{T-(k+1)}(x) & \hspace{-2mm} = \hspace{-2mm} & (1-p_x^d){\cal G}_{T-k}(x) + k(1-p_x^d){\cal U}(x) \\
   &\hspace{-2mm} \hspace{-2mm}& +  p_d \bigg[ {\cal H}_{T-k}(x,{\bf a}^\Delta) 
 +k {\cal V}(x,{\bf a}^\Delta)\bigg], \\
    && \hspace{-25mm} \le \   \big[ C_1(k)(||x||_\infty+1) + C_2(k) \big] +k(1+p_x^d)d||x||_\infty, \\
    && \hspace{-12mm} + p_d \big[ C_1(k)(||x||_\infty+1) +C_2(k) \big], \\
   && \hspace{-25mm} = \ \big[(1+p_d)C_1(k) +kd(1+p_x^d) \big]||x||_\infty  \hspace{-1.4mm}  +  (1+p_d)\big[ C_1(k) + C_2(k) \big],  \\
     && \hspace{-25mm} \le  \ \big[(1+p_d)C_1(k) +2kd \big]||x||_\infty   +  (1+p_d)\big[ C_1(k) + C_2(k) \big],  \\
   && \hspace{-25mm} = \ C_1(k+1)||x||_\infty + C_2(k+1).
\end{eqnarray*}}%
In the above we used \eqref{eqn_C1} and \eqref{eqn_C2}  and then \eqref{eqn_bound} is satisfied.
 Further, since it is a finite horizon problem, the above constants are bounded and since $p \le 1$, the bounds can be independent of $p$.
 \eop


\begin{lemma}
    \label{lem_expect}
  {\it   Given any $x$ and $\a \in {\cal A}_x$, we have
    $$
    \mathbb{E} \left[ \sum_{n \in {\cal N} } H_n' | x, \a \right] = \sum_{n \in {\cal N}} h_n + N + pl(x,\a).
    $$ }
\end{lemma}
\vspace{2mm}
\noindent 
\textbf{Proof: }The above conditional expectation equals,

\vspace{-4mm}
{\small 
\begin{eqnarray*}
    &=& \sum_{n \in \a} \big[ p(g_n+1) + (1-p) (h_n+1) \big] + \sum_{n \notin \a} (h_n+1), \\
    & = & \sum_{n \in {\cal N}} h_n + N + pl(x,\a). \mbox { \hspace{35mm} \eop}
\end{eqnarray*}}

\begin{lemma}
\label{lem_F_G}
{\it  Consider any $x$ and $\a \in \mathcal {A}_x$. For any time $t$ conditioned on $X_t = x$ and ${\cal A}_t = \a$, and with $X'$ representing the quantities at $t+1$, we have the following:
there exists two non-negative-valued bounded functions $\mathcal{U}$ and $\mathcal{V}$  such that the former depends only on $x$ (independent of $t, \a$) while the latter depends on both $x$ and $\a$ (independent of $t$) and one can express the conditional expectation as,

\vspace{-2mm}
 {\small   $$
   \mathbb{E} \bigg[    l(X')   | x,{\bf a} \bigg]  =   (1-p_x^d)\mathcal{U}(x) + p_d   \mathcal{V}(x,\a). 
   $$ 
   where $p_x^d = 1-(1-p)^{N_x^d}$.}
 and $p_d = 1- (1-p)^d$.  Further irrespective of $\a$
$$
|\mathcal{U}(x) |\le     d|| x||_\infty \text{ and } |\mathcal{V}(x,\a) | \le     d|| x||_\infty. 
 $$ 
   }
\end{lemma}
\vspace{1mm}
\noindent
\textbf{Proof: }(i) Let $\mathbb{S}$ be a flag  indicating at least one successful packet transfer. Then, 

\vspace{-2mm}
{\small \begin{eqnarray*}
   \mathbb{E} \bigg[   l(X')| x,\a  \bigg]  
  & = &   (1-p_x^d)   \mathbb{E} \bigg[ l(X')| x,\a ,\mathbb{S} = 0 \bigg]  \\
  && +   p_x^d \mathbb{E} \bigg[  l(X')| x,\a ,\mathbb{S} = 1\bigg].  
\end{eqnarray*}}
Now, it is easy to observe that the state transitions corresponding to first term are independent of the action chosen as $G_{\hat a}' =  1_{\{\mathbb{R}_{\hat a}=0\}} (g_{\hat a}+1)$ and $H_{\hat a}' = h_{\hat a}+1$ for all ${\hat a} \in S_{X'}$ where $\mathbb{R}_{\hat a}$ be the indicator of new packet arrivals at source ${\hat a}$ (observe here that with $\mathbb{S}=0$, $S_X \subset S_{X'}$ and $\mathbb{R}_{\hat a} = 1$ for ${\hat a} \in S_{X'} \setminus S_X$ for almost all $X'$). Thus for some appropriate function ${\cal U}$ of $x$ (alone), the first term equals,
\begin{equation}
 (1-p_x^d)\mathbb{E} \bigg[  l(X')| x,\a ,\mathbb{S} = 0 \bigg]  = (1-p_x^d) {\cal U}(x). \nonumber   
\end{equation}
When conditioned on ${\mathbb S}=1$, it is clear that transitions depend not only on $x$ but also on $\a$. Hence, there exists a function $\mathcal{V}(\cdot)$ of $x$ and $\a$ such that,

\vspace{-3mm}
{\small$$
p_x^d \mathbb{E} \bigg[  l(X')| x,\a ,\mathbb{S} = 1\bigg] = p_d \frac{p_x^d} {p_d}\mathbb{E} \bigg[  l(X')| x,\a ,\mathbb{S} = 1\bigg] = p_d \mathcal{V}(x,\a).
$$}
Now we are left to derive the upper bound on functions ${\cal U}$ and ${\cal V}.$  Consider any source ${\hat a}$ with packet in the new state $X'$. For this source, 
when conditioned on $\mathbb{S}=0$,
the $G_{\hat a}-H_{\hat a}$ term has the following probabilistic description (see \eqref{eqn_h_evol}-\eqref{eqn_trans_prob}), 

\vspace{-2.5mm}
{\small \begin{eqnarray*}
    \big[(G_{\hat a}'-H_{\hat a}')   | x,\a ,\mathbb{S} = 0\big]   
   &= & 
 \begin{cases}
  g_{\hat a}-h_{\hat a}  &  \text{w.p. } (1-q_{\hat a}), \\
  (-1-h_{\hat a}) & \text{w.p. } q_{\hat a}. 
 \end{cases}
\end{eqnarray*}}%
In either case, i.e., almost surely, $(G_{\hat a}'-H_{\hat a}') $ is  upper bounded by $h_{{\hat a}} + 1$  (recall $g_{\hat a} < h_{\hat a}$)
and hence the absolute value   $|l(X')|$ is a.s. upper bounded by $d ||x||_\infty$. Therefore   $|{\cal U}|$ is also   upper bounded by the same quantity.
%
%
%
%
When conditioned on $\mathbb{S}=1$,
the $G_{\hat a}-H_{\hat a}$ terms have the following probabilistic description, and result follows by similar logic,

\vspace{-2mm}

{\small \begin{eqnarray*}
    \big[(G_{\hat a}'-H_{\hat a}')   | x,\a ,\mathbb{S} = 1\big]   \hspace{-30mm} \\
   &= & 
 \begin{cases}
  g_{\hat a}-h_{\hat a}  &  \text{ if } \mathbb{R}_{\hat a} = 0 \text{ and }  {\hat a} \notin \a, \text{ w.p. }(1-q_{\hat a}), \\
  (-1-h_{\hat a}) &  \text{ if } \mathbb{R}_{\hat a} = 1 \text{ and } {\hat a} \notin \a, \text{ w.p. }q_{\hat a}, \\
 (-g_{\hat a}-1) &  \text{ if } \mathbb{R}_{\hat a} = 1 \text{ and }  {\hat a} \in \a, \text{ w.p. }q_{\hat a}.  \ \ \  \ \mbox{ \eop}
 \end{cases}
\end{eqnarray*}}

\end{document}